\documentclass[12pt]{article}
\usepackage{amssymb,amsmath}
\makeatletter

\@addtoreset{equation}{section}
\def\section{\@startsection {section}{1}{\z@}{-2.5ex plus -1ex minus
 -.2ex}{1.3ex plus .2ex}{\large\bf}}
\def\subsection{\@startsection{subsection}{2}{\z@}{-2.25ex plus%
 -1ex minus -.2ex}{0.5ex plus .2ex}{\bf}}

\def\Ad{\mbox{Ad}}
\def\cd{\!\cdot\!}

\def\Poi{ P_3^\uparrow }

\def\LLor{\tilde L_3^\uparrow}
\def\PPoi{\tilde P_3^\uparrow }

\def\bv{{\mbox{\boldmath $v$}}}

\def\ba{{\mbox{\boldmath $a$}}}
\def\bx{{\mbox{\boldmath $x$}}}
\def\by{{\mbox{\boldmath $y$}}}
\def\bj{{\mbox{\boldmath $j$}}}
\def\bk{{\mbox{\boldmath $k$}}}
\def\bp{{\mbox{\boldmath $p$}}}

\def\bq{{\mbox{\boldmath $q$}}}

\newcommand{\ZZ}{\mathbb{Z}}

\newcommand{\RR}{\mathbb{R}}
\newcommand{\CC}{\mathbb{C}}

\def\bea{\begin{eqnarray}}
\def\eea{\end{eqnarray}}
\newtheorem{theorem}{Theorem}[section]

\newtheorem{lemma}[theorem]{Lemma}

\def\bmz{\left(\begin{array}{2,2}}
\def\emz{\end{array}\right)}
\def\bmd{\left(\begin{array}{3,3}}
\def\emd{\end{array}\right)}
\newcommand{\frg}[0]{\mathfrak{g}}
\newcommand{\ve}[0]{{\nu}}
\newcommand{\ad}[0] {\text{ad}}

\newcommand{\mi}[0]{{M_i}}
\newcommand{\ai}[0]{{A_i}}
\newcommand{\bi}[0]{{B_i}}

\newcommand{\mip}[0]{{M'_i}}
\newcommand{\aip}[0]{{A'_i}}
\newcommand{\bip}[0]{{B'_i}}
\newcommand{\mjp}[0]{{M'_j}}
\newcommand{\ajp}[0]{{A'_j}}
\newcommand{\bjp}[0]{{B'_j}}

\newcommand{\cala}[0]{\mathcal{A}}
\newcommand{\calt}[0]{\mathcal{T}}
\newcommand{\cals}[0]{\mathcal{S}}
\newcommand{\calb}[0]{\mathcal{B}}
\newcommand{\cald}[0]{\mathcal{D}}
\newcommand{\calp}[0]{\mathcal{P}}

\begin{document}
\parskip 6pt
\parindent 0pt
\begin{flushright}
HWM-03-2\\
EMPG-03-02\\
gr-qc/0301108
\end{flushright}

\begin{center}
\baselineskip 24 pt
{\Large \bf   Poisson structure and symmetry } \\
 {\Large \bf  in the Chern-Simons formulation of (2+1)-dimensional gravity}

\vspace{1cm}
{\large C.~Meusburger\footnote{\tt  cm@ma.hw.ac.uk}
and    B.~J.~Schroers\footnote{\tt bernd@ma.hw.ac.uk} \\
Department of Mathematics, Heriot-Watt University \\
Edinburgh EH14 4AS, United Kingdom } \\

\vspace{0.5cm}

{26 January 2003}

\end{center}

\begin{abstract}
\baselineskip 12pt
\noindent
In the  formulation of (2+1)-dimensional gravity
 as a Chern-Simons  gauge theory, the phase space is
the moduli space of flat Poincar\'e group connections.
Using the combinatorial approach  developed by Fock and Rosly, we
give an explicit description of the phase space and its
Poisson structure for  the general case of a
 genus $g$ oriented surface with punctures
 representing particles and a
 boundary playing the role of spatial infinity.
We give a physical interpretation and
 explain how the degrees of freedom associated with each
handle and each  particle  can be decoupled.  The
symmetry group of the theory  combines an
action of the mapping class group with asymptotic
Poincar\'e transformations in a non-trivial fashion. We derive the
conserved quantities associated to the latter
and show
 that  the mapping class group of the  surface
acts on the phase
space via Poisson isomorphisms.

\end{abstract}

\section{Introduction}

\baselineskip 15pt

The interest of (2+1)-dimensional gravity (two spatial dimensions, one
time dimension) is that it can serve as a toy model for regular
Einstein gravity in  (3+1) dimensions. In the (2+1)-dimensional
case, the Einstein equations of motion reduce to the requirement that
the spacetime be flat outside the regions where matter is
located. As a result, Einstein gravity in (2+1) dimensions is much simpler
than its (3+1)-dimensional analogue.
The only degrees of
freedom present in (2+1)-dimensional gravity are a finite number of
global degrees of freedom related to
matter and the  topology of the spacetime
manifold.
This gives rise to hope that (2+1)-dimensional gravity will allow
one to study conceptual questions related to the quantisation of
general relativity without being hindered by the technical
difficulties  present in the (3+1)-dimensional case.

As the features of a quantum theory are modelled after their classical
counterparts and should reduce to them in the classical limit,
an important prerequisite of the quantisation of (2+1) gravity is
 the investigation of its phase space. This includes  among
others the parametrisation of the phase space,  the determination of  its
Poisson  structure and the identification of symmetries and conserved
quantities. Starting with the work of   Regge and Nelson
\cite{RN},
these questions have been addressed by  a number of
physicists, using a variety of methods, see \cite{Carlipbook} for
further background. For us, the recent work by Matschull \cite{Matschull1}
which is entirely based on the Einstein formulation of  (2+1) gravity
is going to be a key reference.
However, despite these efforts,
a  complete and explicit description  of
the phase space and its Poisson structure
for arbitrarily many (spinning) particles on a surface with
handles is still missing. In view of our opening paragraph it
is furthermore desirable that such a description  provides
 a starting point for an equally explicit  quantisation of the
theory.

It was first noted in \cite{AT} and  further discussed in
\cite{Witten1} that the
Einstein-Hilbert action in (2+1)
gravity can be written as the Chern-Simons action of an appropriate
gauge group.
 If one adopts  this formulation,
 a large body of mathematical knowledge becomes
available and can be applied to the phase space of (2+1)
gravity. The phase space
of a Chern-Simons theory with gauge group $G$ is the  space of
flat $G$ connections modulo gauge transformations, in the following
referred to as the
moduli space. It inherits a Poisson structure from the
canonical Poisson structure on the space of Chern-Simons gauge
fields. The moduli space of flat $G$ connections and its Poisson
structure has been  investigated extensively in mathematics for the case of
compact semi-simple Lie groups $G$.
Of special interest for this article is the work of Fock and Rosly
\cite{FR}, in which the moduli space arises as a quotient of two finite
dimensional spaces  and the  Poisson structure is given explicitly
in terms of a classical $r$-matrix  \cite{FR}.
This description of the moduli space
was developed further in \cite{AMI},\cite{AMII}, using the
language of Poisson-Lie groups.
Moreover, it is the starting  point for a quantisation
procedure called combinatorial quantisation, carried out and described
in detail  in \cite{AGSI}, \cite{AGSII} and \cite{AS}.

 The aim of this article is to apply these methods
to the Chern-Simons formulation of (2+1) gravity
 in order to give a description of the phase
space  that allows for  a systematic study of its physics and which
is amenable to  quantisation. One advantage of this approach is
its generality.  We are able to
 give an explicit parametrisation of the phase space on a
spacetime manifold of topology
$\mathbb{R}\times S^\infty_{g,n}$, where $S^\infty_{g,n}$ is a
genus $g$ oriented surface with punctures   representing $n$
massive particles (possibly with spin)
and a connected boundary
corresponding  to spatial infinity. In this description, the
mathematical concepts and parameters introduced in \cite{FR} have a
natural physical interpretation. We thus
obtain a mathematically rigorous framework in
which the physics questions can be addressed.

Our article is structured as follows. Sect.~\ref{chernsim} contains an introduction to the Chern-Simons
formulation of (2+1) gravity and its phase space as the
moduli space of flat connections.
 We summarise briefly some mathematical results about the moduli
 space,  in particular the work of Fock and Rosly \cite{FR}
 and Schomerus and Alekseev \cite{AS} underlying our description
of the phase space.

In Sect.~\ref{oneparticle} we apply these results to the phase space
of (2+1) gravity on an open surface $S_{0,1}^\infty$
 with a single massive particle. We extend the description of
the moduli space given by Fock and Rosly to
incorporate a boundary representing spatial infinity. This allows us to
give an explicit parametrisation of the phase space and its Poisson
structure as well as a physical interpretation. Relating  the Poisson
structure of the one particle phase space to the symplectic
structure on  the dual of the universal cover of the (2+1)-dimensional
Poincar\'e group, we show that it generalises the Poisson
structure derived by Matschull and Welling \cite{MW} in
the metric formulation.

Sect.~\ref{phasespace} extends these results to the general case of
 a genus $g$ oriented surface with $n$ punctures
 and a boundary representing spatial infinity. After a discussion of
the boundary condition and asymptotically nontrivial gauge
transformations, we describe the phase space by means of a graph and
derive its Poisson structure. The results are given a physical
 interpretation and related to the theory of
Poisson-Lie groups. Applying the
work of Alekseev and Malkin \cite{AMI},\cite{AMII}, we introduce a set
of normal coordinates that decouple the
contributions of different handles and particles.

In Sect.~\ref{symmetry}, we study the symmetries of the phase
space. Following the work of Giulini \cite{giu1}, we identify the
symmetry group of (2+1)  gravity  in the Chern-Simons formulation
for a spacetime of
topology $\mathbb{R}\times S^\infty_{g,n}$. We determine the
quantities that generate asymptotic symmetries via the Poisson bracket
and interpret them in physical terms. The action of large
diffeomorphisms on the spatial surface $S_{g,n}^\infty$ is
investigated, and we prove that the generators of the (full) mapping
class group act on the phase space as Poisson isomorphisms.

Our final section contains  comments on
the relationship between  our results and other approaches
to (2+1)-dimensional
gravity, our conclusions and an outlook. Facts and
definitions about the universal cover of the (2+1)-dimensional
Poincar\'e group as a Poisson-Lie group are
summarised in the appendix.

\section{The Chern-Simons formulation of (2+1)-dimensional
gravity and its phase space as the moduli space of flat connections}
\label{chernsim}

\subsection{Conventions}

In the following we restrict  attention to (2+1)-dimensional
gravity with vanishing cosmological constant in its formulation
as a Chern-Simons theory on a spacetime $M$ of topology
$M\approx\mathbb{R}\times S^\infty_{g,n}$, where
$S^\infty_{g,n}=S^\infty_g-\{z_1,\ldots,z_n\}$ is a disc with $n$
punctures  and
$g$ handles, i.e. an oriented  surface of genus $g$ with $n$
punctures  and a connected boundary.
The punctures at $z_1,\ldots,z_n$ represent
massive particles, the boundary corresponds to spatial infinity.

 Throughout the paper we use units in which the
speed of light is $1$. Exploiting the fact that in (2+1) gravity
Newton's constant has dimensions of inverse mass, we measure masses
in units of $(8\pi G)^{-1}$. Indices are raised and lowered with the
 (2+1)-dimensional Minkowski metric
$\eta_{ab}=$diag$(1,-1,-1)$, and
Einstein summation convention is employed unless stated otherwise. For
the epsilon tensor we choose the convention
$\epsilon_{012}=1$.

 $L_3^\uparrow$ and $P_3^\uparrow
=L_3^\uparrow\ltimes\mathbb{R}^3$  denote respectively the
(2+1)-dimensional proper ortochronous Lorentz and Poincar\'e group and
$\tilde{L}_3^\uparrow$, $\PPoi=\tilde{L}_3^\uparrow\ltimes
\mathbb{R}^3$ their universal covers. The Lie algebra of the group
$\PPoi$ is $\text{Lie}\;\PPoi=iso(2,1)$ with generators $P_a$,
$J_a$, $a=0,1,2$, and the commutator
 \bea
\label{poinccomm}
 [P_a,P_b] =0,
\quad[J_a,J_b]=\epsilon_{abc} J^c,\quad
[J_a,P_b]=\epsilon_{abc} P^c.
\eea
 The generators
$J_a$, $a=0,1,2$, span the Lie algebra $so(2,1)$ of
 $\tilde L_3^\uparrow$. If we write the
 elements of
 $\PPoi$ as
\bea
(u,\ba)\in\PPoi\qquad\text{with}\qquad u\in\tilde L_3^\uparrow,\;
\ba\in\mathbb{R}^3,\nonumber
\eea the group multiplication in $\PPoi$ is given by
\bea
\label{groupmult}
 (u_1,\ba_1)\cdot(u_2,\ba_2)=(u_1\cdot u_2,\ba_1+\Ad(u_1)\ba_2)
\eea
with $\Ad(u)$ denoting the $L_3^\uparrow$ element associated to $u\in
\tilde L_3^\uparrow$.

In the following we adopt the conventions of
\cite{bamus} for parametrising the  (2+1)-dimensional
Lorentz group and its covers, in particular the
parametrisation via the  exponential map
\bea
\exp:so(2,1)\rightarrow \tilde L_3^\uparrow.
\eea
Note that this map
is neither into nor onto, but that we can nevertheless write
any element $u\in \tilde L_3^\uparrow$  in the form
\bea
\label{lparam}
u=\exp(-2\pi n J_0)\exp(-p^a J_a)\qquad\text{with} \qquad n\in \mathbb{Z}.
\eea
 Combining the parameters $p^a$ into a
three-vector
$\bp=(p^0,p^1,p^2)$, we can characterise elliptic elements of $\tilde
L_3^\uparrow$
by the condition
$\bp^2 = p_a p^a
\in(0,(2\pi)^2)$, parabolic elements by $\bp^2=0$ and hyperbolic
elements by $\bp^2<0$.

This allows us to express the adjoint $\Ad(u)$ of an element
$u\in\tilde L_3^\uparrow$ as
\begin{align}
\label{adjoint}
\Ad(u)_{ab}&=\delta_{ab}+(\delta_{ab}-p_a p_b)\left(\sum_{k=1}^\infty
  \frac{(-1)^k(\bp^2)^{k-1}}{(2k)!}\right)+\epsilon_{abc}p^c
\left(\sum_{k=0}^\infty\frac{(-1)^k (\bp^2)^k}{(2k+1)!}\right)\\
&=\left\{
\begin{array}{ll}
\hat p_a \hat p_b+\cos(\sqrt{\bp^2})
\left(\delta_{ab}- \hat p_a\hat p_b\right)+
\sin(\sqrt{\bp^2})\epsilon_{abc}\hat p^c
& \text{for}\;
u\;\text{elliptic}\\
\delta_{ab}+\epsilon_{abc}p^c+ \frac 1 2 p_a p_b
&  \text{for}\;
u\;\text{parabolic}\\
\hat p_a\hat p_b+\cosh(\sqrt{|\bp^2}|)
\left(\delta_{ab}-\hat p_a\hat p_b\right)+
\sinh(\sqrt{|\bp^2|})\epsilon_{abc}\hat p^c
& \text{for}\;
u\;\text{hyperbolic,}\end{array}\right.\; .\nonumber
\end{align}
where $\hat \bp = \bp/\sqrt{|\bp^2|}$.
Elliptic conjugacy classes in the group $\PPoi$ are characterised by two
restrictions on the parameter three-vectors $\bp$, $\ba$
\begin{align}
\label{massspin}
\bp^2=\mu^2\qquad\bp\cd\bj=\mu s\qquad\text{where}
\qquad\bj:=-\Ad(u^{-1})\ba
\end{align}
with parameters $\mu\in(0,2\pi)$, $s\in\mathbb{R}$.

\subsection{(2+1)-dimensional gravity as a Chern-Simons gauge theory}

 In Einstein's original formulation of general
relativity,  the dynamical variable is a  metric $g$ on $M$. For
the Chern-Simons formulation it is essential  to adopt  Cartan's
point of view, where the theory is formulated in terms of the (non-degenerate)
dreibein of   one-forms $e_a$, $a=0,1,2$, and the spin connection
one-forms $\omega_a$, $a=0,1,2$. The dreibein is related to the
metric via
\bea
\label{metdreib}
\eta^{ab} e_a\otimes e_b \, = g,
\eea
and the  one-forms $\omega_a$ should be thought of as
components of the $\tilde L_3^\uparrow$ connection \bea \omega = \omega_a
J^a. \eea The Cartan formulation is equivalent to
Einstein's metric formulation of (2+1) gravity provided
that the dreibein is invertible.

The vacuum Einstein-Hilbert action
 in (2+1) dimensions can be written in terms of dreibein and spin
connection as
\bea
\label{EHaction}
S_{EH}[\omega, e]= \int_M \,\, e_a \wedge F_\omega^a,
\eea
where $F_\omega^a$ denotes the components of the curvature two-form:
\begin{align}
\label{spincurv}
&F_\omega= d\omega + \frac{1}{2}[\omega,\omega]= F^a_\omega J_a, &
&\;F_\omega^a = d\omega^a  + \frac{1}{2} \epsilon^a_{\;bc}
\omega^b\wedge \omega^c.
\end{align}
Both the connection $\omega_a$ and the
dreibein $e_a$ are  dynamical variables and
varied independently. Variation with respect to the spin
connection yields the requirement that  torsion vanishes:
\bea
\label{notorsion} D_\omega e_a = de_a+  \epsilon_{abc}
\omega^b e^c =0.
\eea
 Variation with respect to $e_a$
yields the vanishing of the curvature tensor:
\bea
\label{Einsteineq}
F_{\omega}=0.
\eea
 In  (2+1) dimensions, this is equivalent
to the Einstein equations in  the
absence of matter.

For the Chern-Simons formulation of gravity,
 dreibein and the spin connection are combined into a
 Cartan connection \cite{Sharpe}. This  is  a  one-form
with values in the Lie algebra $iso(2,1)$
\bea
\label{Cartan} A = \omega_a
J^a + e_a P^a,
\eea
whose  curvature
 \bea
\label{decomp} F = (D_\omega e^a) P_a + (F_\omega^a) J_a
\eea
combines the curvature and the torsion of the spin connection.

The final ingredient needed to establish
the Chern-Simons
formulation is a non-degenerate, invariant
  bilinear form on the Lie algebra $iso(2,1)$
\bea
\label{inprod}
\langle J_a, P_b\rangle = \eta_{ab}, \quad  \langle J_a, J_b\rangle
= \langle P_a,P_b\rangle = 0.
\eea
Then the  Chern-Simons action  for the
connection
$A$ on  $M$ is
\bea
\label{CSaction}
S_{CS}[A] =\frac{1}{2} \int_M \langle A\wedge  dA\rangle
 +\frac{2}{3}\langle A \wedge A \wedge A\rangle.
\eea

A short calculation shows that this is equal to the Einstein-Hilbert
 action (\ref{EHaction}). Moreover, the equation of motion found
by varying the action with respect to $A$ is
\bea
\label{flat}
F=0.
\eea
Using the decomposition (\ref{decomp}) we thus reproduce
the condition of vanishing torsion and the three-dimensional
Einstein equations, as required. This shows that there is a one-to-one correspondence
of solutions of Einstein's equations and flat Chern-Simons gauge
fields with non-degenerate dreibein.

 For a spacetime of the form $S_g^\infty -\{z_1,\ldots,z_n\}$
with punctures  corresponding to massive particles, an additional
condition has to be given regarding the behaviour of the curvature
tensor at the punctures. The inclusion of massive particles with spin
into (2+1)-dimensional gravity in its Chern-Simons formulation has
been investigated by several authors. As explained in
\cite{Carlipscat},  the curvature tensor develops $\delta$-function
 singularities at the positions of the particles. For a single
 particle of
mass $\mu$ and spin $s$
 at rest at $z_1$ the curvature tensor is given by
\bea
\label{parteq}
F(z) +\left(\mu J_0 + s P_0\right) \delta(z-z_1)=0
\eea
As a consequence, the holonomy for an infinitesimal circle
 surrounding the particle
is
\bea
\label{element0}
h_0=(e^{-\mu J_0},-(s,0,0)^t).
\eea
 As we will show in Sect.~\ref{oneparticle},  the holonomy for a
 general loop around the particle is related
to $h_0$ via conjugation with a $\PPoi$-element. It is an element of a
fixed elliptic $\PPoi$-conjugacy class parametrised by the particle's
mass $\mu$ and spin $s$ as in \eqref{massspin}.

\subsection{The phase space of Chern-Simons theory}
In a Chern-Simons theory with gauge group $G$ on manifold
$M=\mathbb{R}\times S$, where $S$ is a two-dimensional surface with
connected boundary, two gauge connections
$A$ and $A'$ describe the same physical state if they are related
by a  gauge transformation
\bea
\label{gaugetransf}
A'= \gamma A \gamma^{-1} +\gamma d\gamma^{-1}
\eea
where $\gamma$ is a $G$-valued  function on the surface $S$.
 It must satisfy an appropriate fall-off
condition compatible with the conditions imposed on the gauge
connections at the boundary. The phase space of a
Chern-Simons theory with gauge group $G$ on a manifold
$M$ is the space of all physically distinct
solutions of the equations of motion \eqref{flat} subject to
conditions of the form
\eqref{parteq} at the punctures. It is the moduli
space of flat $G$ connections on $S$ modulo gauge
transformations  \eqref{gaugetransf}.
It inherits a Poisson structure
from the canonical symplectic structure on the space of gauge connections
\cite{AB}.
The properties of the moduli space have
been investigated extensively in mathematics for the case of compact
gauge groups $G$. In particular, it has
been shown that it is finite dimensional. A review of the mathematical
results and further references are given in the book \cite{Atiyah}.

\subsubsection*{The Fock-Rosly description of the moduli space}

Fock and Rosly gave a description of the moduli space of flat
$G$ connections on a closed, oriented   surface $S$ with
punctures by means of a graph embedded into the surface
\cite{FR}, see\cite{audin} for a pedagogical account.
The underlying idea is similar to lattice gauge theory, but - due to
the absence of local degrees of freedom -  the physical content of
the theory is captured entirely by a sufficiently
refined\footnote{As we use only sufficiently refined graphs in
this article, we will not explain this concept further.} graph without
the need to take a continuum limit. According to Fock and Rosly, the
moduli space and its Poisson structure on an oriented,
 surface with punctures  can be uniquely characterised by a
{\em ciliated fat graph} $\Gamma$. This is a set of $N_\Gamma$
vertices and $I_\Gamma$ oriented edges connecting the vertices
together with a linear ordering of the incident edges at each vertex.
If such a graph is embedded into the surface $S$, the orientation
 induces a cyclic ordering of the incident edges
at each vertex and makes the graph a {\em fat graph}. The surface
can be completely reconstructed from a sufficiently refined fat
graph.
The reconstruction of the surface and the Poisson structure
on the moduli space requires a   {\em ciliated fat graph}.
It can be obtained from a fat graph embedded into the
surface by adding a cilium at each vertex in order to
separate the incident edges of minimum and maximum order.
Given a smooth $G$ connection on the surface $S$ and a ciliated fat graph
$\Gamma$ embedded into it, parallel transport along the
edges of the graph assigns an element $A_i$ of the gauge
group to each oriented edge $i$ and thus induces
a {\em graph connection}: a map from the set of oriented
edges into the direct product of $I_\Gamma$ copies of the
gauge group $G$. Flat connections on the surface induce
{\em flat graph connections}, for which the ordered product
of the group elements assigned to edges around a face of
the graph is trivial if the face does not contain any
punctures. Similarly, gauge
transformations \eqref{gaugetransf} induce
transformations of the group elements associated
to each edge: The group element $A_i$ assigned to the oriented
edge $i$ by parallel transport transforms according to
\bea
\label{graphgauge}
A_i\rightarrow \gamma(z([i^\vee]))A_i \gamma^{-1}(z([i])),
\eea
where $z([i])$ denotes the position of the vertex edge $i$ points to
and $z([i^\vee])$ the position of the vertex at which it starts.
This defines a {\em graph gauge transformation}, a map from the
set of vertices into the direct product of $N_\Gamma$ copies of
the gauge group. As explained in \cite{FR},  for any sufficiently refined
graph $\Gamma$  the moduli space on $S$ is isomorphic the quotient
of the space $\mathcal{A}_\Gamma$ of flat graph connections modulo
graph gauge transformations $\mathcal{G}_\Gamma$.
\begin{theorem} The moduli space $\mathcal{M}$ on a closed, oriented
 surface $S$ with punctures is isomorphic to the
quotient of the space of graph connections modulo graph
gauge transformations for any sufficiently refined fat graph $\Gamma$
\bea
\mathcal{M}\cong\mathcal{A}_\Gamma/\mathcal{G}_\Gamma.\nonumber
\eea
\end{theorem}
This characterises the moduli space in terms of
two {\em finite} dimensional spaces. The essential advantage of
the approach by Fock and Rosly is that it allows one  to express the
Poisson structure on the moduli space by a Poisson structure defined
on the space $\mathcal{A}_\Gamma$ of graph connections rather than
the Poisson structure on the (infinite dimensional) space of gauge
connections. The construction of this Poisson structure involves
the assignment of a classical $r$-matrix for the Lie algebra
$\mathfrak{g}$ to each vertex of the graph. This is an element
$r\in\frg\otimes\frg$ which
satisfies the classical Yang-Baxter equation \eqref{cybe}
and whose symmetric part is equal to the
tensor representing a non-degenerate invariant bilinear form  on $\frg$;
details are given in the appendix.
\begin{theorem} (Fock, Rosly)
\label{frpoiss}
\begin{enumerate}
\item Let $\Gamma$ be a sufficiently refined ciliated fat graph. Assign
 a $r$-matrix to each vertex $\ve$, and let $r^{\alpha\beta}(\ve)$ be its
 components with respect to a basis $\{X_\alpha\}$,
$\alpha=1, ..., $dim$\,\frg$ of $\frg$.
Then the following bivector defines a Poisson
structure on the space of graph connections $\mathcal{A}_\Gamma$
\bea
\label{frbivect}
B_{FR}&=&\sum_{\text{vertices}\;\ve=1}^{N_\Gamma}\bigg(
\frac{1}{2}\sum_{i\in\ve}r^{\alpha\beta}(\ve)R_\alpha^i\wedge
    R_\beta^i
+\frac{1}{2}\sum_{i^\vee\in\ve}r^{\alpha\beta}(\ve)
L_\alpha^i\wedge L_\beta^i\nonumber \\
&+&\sum_{i,j\in\ve,i<j}r^{\alpha\beta}(\ve)R_\alpha^i\wedge
R_\beta^j
+\sum_{i,j^\vee\in\ve,i<j^\vee}r^{\alpha\beta}(\ve)R_\alpha^i\wedge
L_\beta^j\nonumber\\
&+&\sum_{i^\vee,j\in\ve,i^\vee<j}r^{\alpha\beta}(\ve)L_\alpha^i\wedge
R_\beta^j\nonumber
+\sum_{i^\vee,j^\vee\in\ve,i^\vee<j^\vee}r^{\alpha\beta}(\ve)
L_\alpha^i\wedge
L_\beta^j\\
&+&\sum_{i,i^\vee\in\ve}r^{\alpha\beta}(\ve)R_\alpha^i\wedge
L_\beta^i\bigg)\;,
\eea
where $i\in\ve$ denotes an edge pointing towards vertex
$\ve$, $i^\vee\in\ve$ an edge starting at vertex $\ve$. $R^i_\alpha$
and $L^i_\alpha$, respectively, are the left and right invariant
vector
fields associated to edge $i$ with respect to the basis $X_\alpha$ of
$\mathfrak{g}$ and $>$, $<$ refer to the ordering of the incident
edges at each vertex. The convention $i<i^\vee$ is chosen for each
edge starting and ending at the same vertex.

\item Let  $\mathcal{G}_\Gamma\cong G^{N_\Gamma}$
  be equipped with the $N_\Gamma$-fold direct product of the Poisson structure on
the group $G$ that is defined by means of a $r$-matrix as described in the
appendix. The elements of the group
$\mathcal{G}_\Gamma$ of graph gauge transformations act as
Poisson maps $\mathcal{G}_\Gamma\times \mathcal{A}_\Gamma\rightarrow \mathcal{A}_\Gamma$
with respect to the direct product Poisson structure on
$\mathcal{G}_\Gamma\times \mathcal{A}_\Gamma$ and the Poisson structure on
$\mathcal{A}_\Gamma$ defined by \eqref{frbivect}.
\item As graph gauge transformations are Poisson maps,
the Poisson structure defined by \eqref{frbivect} induces a
Poisson structure on the quotient $\mathcal{M}$. It is independent
of the graph and isomorphic to the Poisson structure induced by
the canonical symplectic structure on the space
of gauge connections.
\end{enumerate}
\end{theorem}
The Poisson bivector \eqref{frbivect} can be decomposed
into a part tangential to the gauge orbits and a part transversal to them:
\begin{align}
\label{frbivect2}
B_{FR}=&\bigg(\sum_{\text{vertices}\;\ve=1}^{N_\Gamma}r^{\alpha\beta}_{(a)}
(\ve)
\big(\sum_{i\in\ve}R_\alpha^i+\sum_{i^\vee\in\ve} L_\alpha^i\big)\otimes
\big(\sum_{j\in\ve}R_\beta^j+\sum_{j^\vee\in\ve}
L_\beta^j\big)\bigg)\\
+&\bigg(\sum_{{\text vertices}\;\ve=1}^{N_\Gamma}
t^{\alpha\beta}\sum_{i,j\in\ve,i<j}R_\alpha^i\wedge
R_\beta^j
+\sum_{i,j^\vee\in\ve,i<j^\vee}R_\alpha^i\wedge
L_\beta^j
+\sum_{i^\vee,j\in\ve,i^\vee<j}L_\alpha^i\wedge
R_\beta^j\nonumber
\\
&\qquad+\sum_{i^\vee,j^\vee\in\ve,i^\vee<j^\vee}L_\alpha^i\wedge
L_\beta^j\nonumber+\sum_{i,i^\vee\in\ve}R_\alpha^i\wedge
L_\beta^i\bigg)\;.
\end{align}
 The first line in formula \eqref{frbivect2} depends only on the
 antisymmetric parts
\bea
 r_{(a)}^{\alpha\beta}(\ve)=
\tfrac{1}{2}(r^{\alpha\beta}(\ve)-r^{\beta\alpha}(\ve))\eea
 of the $r$-matrices assigned to each vertex and is tangential to the
 gauge orbits. The second part of the bivector and with it the Poisson
 structure on the moduli space depends only on the symmetric part
 common to all $r$-matrices, the components
\bea
 t^{\alpha\beta}=\tfrac{1}{2}(r^{\alpha\beta}(\ve)+r^{\beta\alpha}(\ve))\eea
 of the matrix representing the bilinear form on the Lie algebra
$\mathfrak{g}$. In particular, this implies that the Poisson structure
on the moduli space is not affected by the choice of the $r$-matrix at
each vertex.

\subsubsection*{The description using the fundamental group}
Alekseev, Grosse and Schomerus specialised this description of the moduli
space to the simplest graph that can be used to characterise a closed
 surface with punctures: a choice of generators of its fundamental
group \cite{AGSI}, \cite{AGSII} \cite{AS}. For a closed
surface $S$ of genus $g\geq 0$ with $n\geq 0$ punctures, the
fundamental group
$\pi_1(z_0,S)$ with respect to a basepoint $z_0$
 is generated by $2g+n$ curves starting and ending at $z_0$,
two curves $a_j$, $b_j$, $j=1,\ldots,g$, around each handle and a loop
$m_i$, $i=1,\ldots,n$, around each puncture (see Fig.~1).
\vbox{
\vskip .3in
\input epsf
\epsfxsize=12truecm
\centerline{
\epsfbox{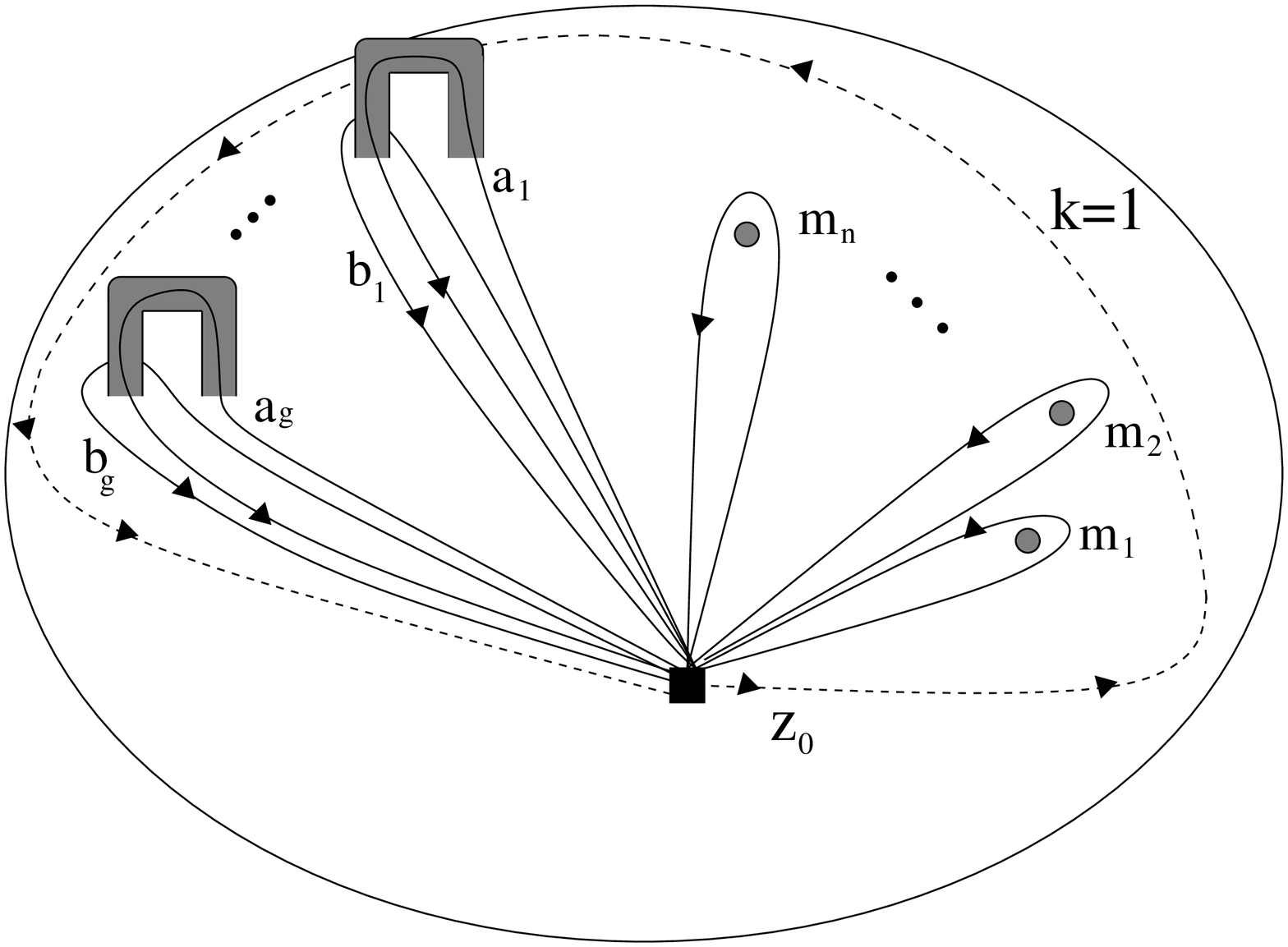}
}
\bigskip
{
\centerline{\bf Fig.~1 }
}
\centerline{\footnotesize The generators of the fundamental group on a genus g
   surface with n punctures }
\bigskip
}
 Its generators obey a
single relation
\bea
\label{relation}
k=[b_g,a_g^{-1}]\cdot\ldots\cdot [b_1,a_1^{-1}]\cdot m_n\cdot\ldots\cdot m_1=1,
\eea
where $[\,,\,]$ is the group commutator $[x,y]=xyx^{-1}y^{-1}$. Via
the holonomy, a graph connection assigns an element of the gauge group
$G$ to each of the generators. Due to equation \eqref{parteq}, the
holonomies $M_i$ of the loops $m_i$ associated to the punctures must
be elements of a fixed conjugacy classes $\mathcal{C}_i$, whereas
there is no such constraint for the group elements $A_j$, $B_j$
associated to the curves around the handles. Taking this into account,
the space of graph connections is given as
\bea
\mathcal{A}_{\pi_1}&=&\{\phi\in\text{Hom}
(\pi_1(z_\infty,S_{g,n}),G)\;|\;M_i=\phi(m_i)\in\mathcal{C}_i, i=1,\ldots,n\}\\
&\cong&\{(M_1,\ldots,B_g)\in G^{2g+n} \;|\; [B_g,A_g^{-1}]
\cdot\ldots\cdot [B_1,A_1^{-1}]\cdot M_n\cdot\ldots\cdot M_1=1,\nonumber\\
&\;&\qquad\qquad\qquad\qquad\qquad\qquad \qquad\qquad
\qquad\qquad M_i\in\mathcal{C}_i, i=1,\ldots,n\}.\nonumber
\eea As the graph defined by the fundamental group has only
one vertex, the group of graph gauge transformations is
the gauge group $\mathcal{G}_{\pi_1}=G$, acting on the holonomies by
global conjugation,
and the moduli space is the quotient
\bea
\label{modquotient}
\mathcal{M}=\mathcal{A}_{\pi_1}/G.
\eea  The Poisson bivector \eqref{frbivect}
defines a Poisson structure on the space $A_{\pi_1}$ of graph
connections that induces a Poisson structure equivalent to the
canonical Poisson structure on the moduli space.

\subsection{The phase space of (2+1) gravity as
  the moduli space of flat $\PPoi$ connections}

 The material of the
previous two subsections allows us to
formulate in precise, technical terms
 the
questions  we have to address
when  applying the  Fock-Rosly  description
of the moduli space
to the Chern-Simons
formulation of (2+1)-dimensional gravity.

The first ingredient required to implement
the Fock-Rosly description is a classical $r$-matrix for the
Lie algebra $iso(2,1)$ of the Poincar\'e group. As we shall see in the
next section, the required $r$-matrix corresponds to the one
given in \cite{schroers}
for the Euclidean case and can  easily be adapted to the
Lorentzian setting.

Second, we have to choose
the gauge group $G$.  The Chern-Simons action
only depend on
the Lie algebra $iso(2,1)$, and {\it a priori} we are free
to consider the Poincar\'e group or one of its covers as the
gauge group. In this  paper we take
the gauge group  $G$ to be  the universal cover $\PPoi$, as this leads to
 various technical simplifications.

Since essentially all mathematical work on the moduli space
of flat connections has been carried out in the context of
compact gauge groups one might worry about complications caused by
working with the  non-compact  group $\PPoi$. However, the
difficulties associated to non-compact gauge groups arise when
taking quotients by an action of the gauge group. It turns
out that in our application to (2+1) gravity, such quotients never
arise. This is due to the fact that we work with a surface with
boundary and thus closely related to the next challenge,
namely the incorporation of a  surface with boundary in
the Fock-Rosly formalism. The details of how this is done are the
subject of the next section. However, one consequence is that
we do not need to take the
quotient (\ref{modquotient}).

The final issue to be investigated
is that of gauge invariance and symmetry.
The
Chern-Simons formulation of (2+1)-dimensional gravity is invariant
under Chern-Simons gauge transformations as well as under
diffeomorphisms.
However, large diffeomorphisms (not connected to the identity)
and asymptotically non-trivial gauge transformations do not
relate physically equivalent configurations. They combine in a rather
subtle way to
form the symmetry group of the theory.

\section{The phase space for a single massive particle}
\label{oneparticle}

\subsection{The open spacetime containing a single particle}
We begin our investigation of  the phase space of (2+1)-dimensional
gravity with the case of an open universe of topology
$\mathbb{R}\times S_{0,1}^\infty$, containing a single massive
particle. In the metric formalism, a spacetime with a particle of mass
$\mu$  and spin $s$
that is at rest at the origin, is described by the conical metric
\cite{djt}
\begin{align}
\label{conem}
&ds^2=(dt+s\,d\varphi)^2-\frac{1}{(1-\tfrac{\mu}{2\pi})^2}dr^2-r^2
d\varphi^2 &
&\text{with}\;r\in(0,\infty),\, t\in\mathbb{R},\,\varphi\in[0,2\pi),
\end{align}
where the mass of the particle is restricted to the interval
$\mu\in(0,2\pi)$. Deficit angle $\Delta\varphi$ and time shift
$\Delta t$ of the cone are given by
\bea\label{deficit}
\Delta\varphi=\mu\qquad\text{and}\qquad\Delta t=2\pi s.
\eea
A
dreibein and spin connection leading to this metric via
\eqref{metdreib} are
\begin{align}
\label{dreib}
&e^0=dt+s\,d\varphi & &\omega^0=\tfrac{\mu}{2\pi}d\varphi\\
&e^1=\frac{1}{1-\tfrac{\mu}{2\pi}}\cos\varphi \,dr-r\sin\varphi\,
d\varphi & &\omega^1=0\nonumber\\
&e^2=\frac{1}{1-\tfrac{\mu}{2\pi}}\sin\varphi \,dr+r\cos\varphi \,
d\varphi & &\omega^2=0\nonumber.
\end{align} With the conventions given in Sect.~2,
they can be combined into a Chern-Simons gauge field
\bea
\label{gaugefield}
A_\infty=\omega^aJ_a+e^aP_a.
\eea

As the spatial surface $S^\infty_{0,1}$ has  a
boundary representing  spatial infinity,
we must impose an appropriate boundary condition on the gauge field
and restrict  to  gauge transformations  which are
compatible with this condition. We derive the boundary condition in
the Chern-Simons formulation from a corresponding boundary condition
in the metric formalism. In a reference frame where the massive
particle is at rest at the origin, the particle's centre of
mass frame) the boundary condition in the
metric formalism is the (trivial) condition that the metric be conical
 of the form \eqref{conem}. The gauge transformations  compatible with
this condition must reduce to spatial rotations and translations in the time
direction outside an open region containing the particle.

In the Chern-Simons formulation of (2+1)-dimensional gravity,
the metric \eqref{conem} corresponds to a gauge field
\eqref{gaugefield}
with $e$, $\omega$ given by \eqref{dreib}.
 The admissible gauge transformations
are asymptotically constant Chern-Simons gauge transformations.
Spatial rotations are implemented by
Chern-Simons gauge transformations \eqref{gaugetransf} of $A_\infty$,
 where $\gamma$ is  constant and takes the value
\bea
\label{spacerot}
g=(\exp(-\theta J_0),0)\nonumber
\eea
with $\theta$ constant outside an open region containing the particle.
For time translations we take
\bea
\label{timetrans}
g=(1,(q_0,0,0)^t)\nonumber
\eea
with $q_0$ constant outside an open region containing the particle. These
asymptotically nontrivial transformations differ in their physical
interpretation from regular gauge transformations which vanish outside
a  region containing the particle. They are not related to gauge
degrees of freedom but physically meaningful transformations acting on
the phase space. We return to this question
in the next section when we consider a
universe containing an arbitrary number of particles.

Note, however, that in
the Chern-Simons formulation it is not necessary to limit
oneself to the particle's centre of mass frame.
In order to study the dynamics of a single particle
we should admit general Poincar\'e
transformations with respect to the centre of mass  frame.
Then the admissible transformations are Chern-Simons
transformations of the form \eqref{gaugetransf} with $d\gamma=0$
outside an open region containing the particle. The corresponding
boundary condition on the gauge field is the requirement that the
gauge field be obtained from $A_\infty$ \eqref{gaugefield}
by  an asymptotically constant gauge
transformation  $g=(\Lambda,\bq)$. i.e. that it be  of the form
$ \Ad(g) A_\infty $.

\subsection{Phase space and Poisson structure}

 The simplest graph describing the open  spacetime containing a single
 consists of a single vertex $z_\infty$ at the boundary  and a loop
 around the particle. The loop can be built up from an edge connecting
 vertex and particle and an (infinitesimal) circle around the particle
  as pictured in Fig.~2.

\vbox{
\vskip .3in
\input epsf
\epsfxsize=12truecm
\centerline{
\epsfbox{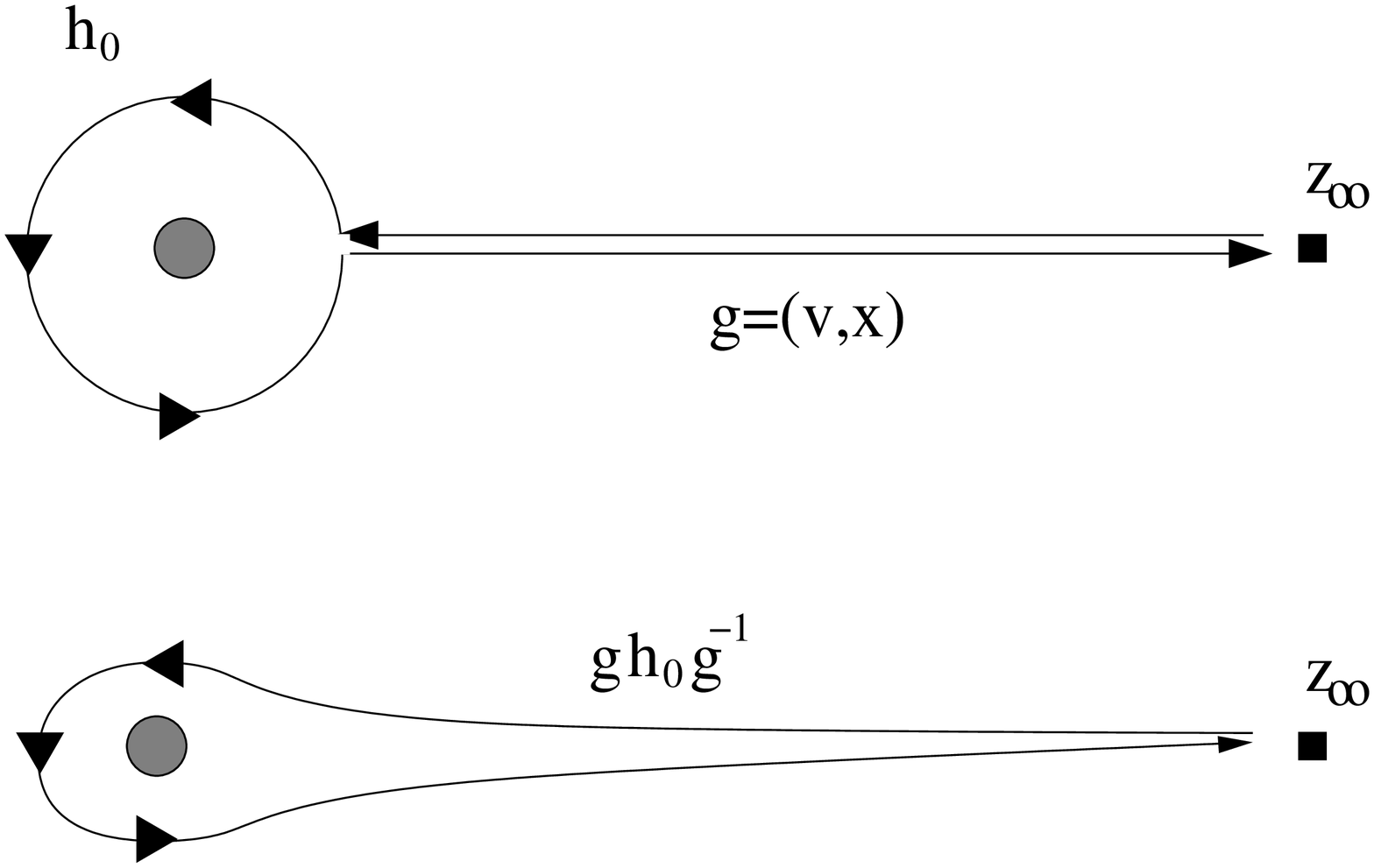}
}
\bigskip
{
\centerline{\bf Fig.~2 }
}
\centerline{\footnotesize The holonomy of a loop around a puncture}
\bigskip}

Using expression \eqref{dreib}, we calculate that
 the  holonomy around the
infinitesimal circle is the element $h_0$ given
by \eqref{element0}. If we write  $g=(v,\bx)$ for
the group element obtained by parallel transport along the edge
connecting vertex and particle, the holonomy $h$ of the loop
starting and ending at $z_\infty$  is
\bea
\label{element}
h=g h_0 g^{-1}.
\eea
Defining
\bea
\label{coord}
p^a J_a=\mu \Ad( v )J_0
\eea
and
\bea
\label{umom}
u=v\exp(-\mu J_0)v^{-1}=\exp(-p^aJ_a)
\eea
 we can also write the holonomy as
\bea \label{holonomy}h=(u,\ba)=(u,-\Ad{(u)}\bj),
\eea
where
\bea
\label{angmom}
\bj=(1-\Ad(u^{-1}))\bx + s\hat \bp.
\eea
The fact that the holonomy is an element of a fixed
$\PPoi$-conjugacy class determined by mass $\mu$ and
spin $s$ of the particles results in constraints on the
parameters $p_a$ and $j_a$
\bea
\label{constraints}
\bp^2=p_a p^a = \mu^2\qquad \bp\cd\bj=p_a j^a =  \mu s.
\eea
Graph gauge transformations arise from asymptotically nontrivial
gauge transformations \eqref{gaugetransf} and act on the
graph by conjugating the
holonomy. The moduli space on the surface would be obtained by dividing
this transformations out. However, as asymptotically nontrivial
transformations are physically meaningful, we do not want to do this.
The extended phase space is then simply the group $\PPoi$ parametrised
by the parameter three-vectors $\bp$ and $\bj$.
The physical phase space is obtained from it by imposing the
constraints \eqref{constraints}, which selects
the $\PPoi$-conjugacy class given by \eqref{element}.

The Poisson structure on the phase space is defined by the
Fock-Rosly bivector \eqref{frbivect}. For the graph introduced
 above it is given by
\bea
\label{onepartbivect}
B_{FR}=\tfrac{1}{2} r^{\alpha\beta}R_\alpha\wedge
R_\beta+\tfrac{1}{2}r^{\alpha\beta}L_\alpha\wedge
L_\beta+r^{\alpha\beta}R_\alpha\wedge L_\beta.
\eea
As we do not divide by graph gauge transformations,
the choice of the $r$-matrix is no longer
arbitrary
and different $r$-matrices lead to different Poisson structures on the
phase
space. The correct choice of the $r$-matrix has to be determined
from physical considerations: the components of the vectors $\bp$,
$\bj$
should have the Poisson brackets expected for momentum and
angular momentum three-vector of a free relativistic particle and act as
infinitesimal generators
of asymptotic Poincar\'e transformations.
As it turns out, this requires the $r$-matrix
\bea
\label{rmatrix}
r=P_a\otimes J^a.
\eea The right and left invariant vector fields on $\PPoi$ are given by
\begin{align}
\label{vecfields}
P_a^R f(h)&:=\frac{d}{dt}|_{t=0}f(he^{tP_a})= -\frac{\partial f}
{\partial j^a}(h)\\
P_a^Lf(h)&:=\frac{d}{dt}|_{t=0}f(e^{-tP_a}h)=
\Ad{(u)}_{ab}\frac{\partial f}{\partial j_b}(h)\nonumber\\
J_a^R f(h)&:=\frac{d}{dt}|_{t=0}f(he^{tJ_a})=
\big(\frac{1}{\Ad(u)-1}\big)_{ab}\epsilon^b_{\;cd}p^c
\frac{\partial f}{\partial p^d}(h)-\epsilon_{abc}j^b\frac{\partial f}
{\partial j_c}(h)\nonumber\\
J_a^L f(h)&:=\frac{d}{dt}|_{t=0}f(e^{-tJ_a}h)=\big(\frac{\Ad(u)}
{1-\Ad(u)}\big)_{ab}\epsilon^b_{\;cd}p^c\frac{\partial f}
{\partial p^d}(h)\nonumber,
\end{align}
where all functions are evaluated at $h$ parametrised in terms of
$\bp$ and $\bj$ as in \eqref{holonomy}\footnote{Note that the
analogous relation between the Euclidean group element $g$ and $
\bp$ and $\bj$  is stated incorrectly at one point in \cite{schroers}}.
  The  expression
$\bigl(\Ad(u)/(1-\Ad(u))\big)_{ab}\epsilon^b_{\;cd}p^c$ in
\eqref{vecfields} is defined properly by
\bea
\label{tdef}
\big(\frac{\Ad(u)}{1-\Ad(u)}\big)_{ab}\epsilon^b_{\;cd}p^c =
\Pi+\frac{\tfrac{\mu}{2}}{\sin(\tfrac{\mu}{2})}\exp(-\frac{1}{2}p^aJ_a)(1-\Pi),
\eea
where $\Pi$ denotes the projector onto the direction of the momentum
$\Pi_a^{\,\;b}=\hat{p}_a\hat p^b$. Note that there is still a
coordinate singularity in the limit
$\mu\rightarrow 2\pi$, where
the coordinates \eqref{coord}  no longer provide a  good parametrisation of the
holonomy $h$.

Inserting the vector fields \eqref{vecfields} and the
$r$-matrix into  equations \eqref{onepartbivect}, we
obtain the Poisson structure in terms of the parameters
$p_a$ and $j_a$, $a=0,1,2$,
\bea
\label{kiri}
\{j_a,j_b\}=-\epsilon_{abc}j^c \qquad \{j_a,p_b\}=
-\epsilon_{abc}p^c \qquad\{p_a,p_b\}=0.
\eea
These equations together with the constraints
\eqref{constraints} suggest  an interpretation of the
parameters $p_a$ as the components of the particle's three-momentum,
$p^0$ as its energy and the spatial components $p^1,p^2$ as its
momentum. Similarly, the vector $\bj$, in the
following referred to as angular momentum
three-vector, contains the particle's angular momentum $j^0$ and the
spatial components $j_1$, $j_2$ associated to the Lorentz boosts.
It can be seen immediately that the constraints \eqref{constraints}
are Casimir functions of this Poisson structure.
Their values parametrise its symplectic leaves, the conjugacy
classes in $\PPoi$.

\subsection{Physical Interpretation}

The components of the graph used to construct the Poisson structure
 can be given a physical interpretation as follows.
The vertex $z_\infty$
 represents an observer.
The $\PPoi$-element $g=(v,\bx)$ assigned to the edge
connecting it with the particle gives the Poincar\'e transformation
relating the observer's reference frame to  centre of mass  frame
of the particle. The vector $\bx$ determines the translation in time
and the  position, $v$ a spatial rotation and/or  Lorentz boosts.

The non-standard relation (\ref{angmom})
between the angular momentum three-vector $\bj$, the momentum $\bp$
 and the position three-vector $\bx$ was
 first discussed in detail for the case of spin-less particles in  \cite{MW}
, where it was derived
 entirely in terms of the metric
formulation of (2+1) gravity. In that paper the authors also
 pointed out that it
leads to non-standard commutation relations
of the position coordinates. In order to
understand this non-commutativity from our point of view,
recall that the
 usual (non-gravitational)
formula for the angular momentum three-vector of a free
relativistic particle  is obtained using the co-adjoint
action (instead of the conjugation action above)
\bea
p^aJ_a+k^aP_a=g(\mu J_0 + sP_0)g^{-1},
\eea
leading to the familiar formula
 for the  angular momentum three-vector $\bk$
\bea
\label{angmomm}
\bk=\bx\wedge \bp +  s \hat \bp.
\eea
Note that the formula (\ref{angmom}) approaches (\ref{angmomm})
in the limit $\mu\rightarrow 0$. More generally
the relationship between $\bj$ and $\bk$ is given
by the Lorentz transformation
\bea
\label{Ttrans}
T(\bp)=\frac{\exp(p^aJ_a)-1}{p^aJ_a}.
\eea
Since this transformation, which is the inverse of \eqref{tdef},
plays an important role in the rest of
the paper, we express it more concretely as a power series
valid for any value of the momentum three-vector $\bp$
\bea
T(\bp)=\sum_{n=0}^\infty \frac{1}{(n+1)!}( p^aJ_a)^n.
\eea
 In the case
at hand, where $\bp$ is a time-like vector of length $\mu$, this yields

\bea
T(\bp)=\Pi  + \frac{\sin\left(\frac{\mu}{2}\right)}{\frac{\mu}{2}}
\exp\left(\frac{1}{2}p^aJ_a\right)(1-\Pi),
\eea
where we again used the projector defined after (\ref{tdef}).
Using  the adjoint or vector representation
of the $J_a$, i.e. $(J_a)_b^{\;\,c}=-\epsilon_{ab}^{\quad c}$, one checks that
\bea
j_a=T(\bp)_a^{\;\, b} k_b.
\eea
Note that in terms of  the transformed position vector
\bea
\label{postrans}
q_a=T(\bp)_a^{\;\,b} x_b,
\eea
the expression for (\ref{angmom}) takes on the ``non-gravitational''
form
\bea
\label{angmommm}
\bj=\bq\wedge \bp +  s \hat \bp.
\eea
This formula
 goes some way towards explaining the non-standard commutation
relations of the position coordinates $x_a$.
They differ from those of the $q_a$ (in terms
of which $j_a$ has the familiar form)  because $\bq$ and $\bx$
are related by the $\bp$-dependent transformation $T$.

\subsection{The Relation to the Dual of the Poisson Lie group $\PPoi$}

For the case of vanishing spin $s$, our description of the
one-particle phase space agrees with the description derived by
Matschull and Welling
\cite{MW}, \cite{Matschull1} in a completely independent way and
generalises their results to the case of arbitrary spin.
In order to see this, we have to describe the Poisson structure
in terms of a symplectic form rather than a Poisson bivector.
It follows from the general results in  \cite{AMI}
that the Poisson bivector \eqref{onepartbivect}
   gives  the Poisson structure
of the dual  Poisson-Lie group  $(\PPoi)^*$.
In the appendix we show that, as a group, the dual is the direct
product
$(\PPoi)^*\cong \LLor\times \RR^3$.
 Elements $h=(u,\ba)\in \PPoi$ can be factorised uniquely in
the form \eqref{factoring} $h=h_+h_-^{-1}$,  where $h_+=(u,0)$ and
$h_-=(1,-\Ad(u^{-1})\ba)=
(1,\bj)$ can naturally be thought of as elements of $(\PPoi)^*$.
The symplectic leaves are precisely the conjugacy classes of $h$.
 The symplectic form on those symplectic
leaves can be written in terms of the parametrisation \eqref{element}
as  \cite{AMI}
\bea
\Omega=\frac 1 2 \langle (dh_+ h_+^{-1} - dh_-h_-^{-1})\wedge
dg g^{-1}\rangle.
\eea
Explicitly, we find  that $dh_+ h_+^{-1}$ is
the right-invariant one-form on $\LLor$
\bea
dh_+ h_+^{-1}=du u^{-1}
\eea
with values in the Lie algebra $so(2,1)$,
and
\bea
dh_-h_-^{-1}=dj_a P^a=dx_a \left(1-\Ad(u^{-1})\right) P^a +
\Ad(u^{-1})[duu^{-1},x_a P^a]
+s\,\,d\hat p_a\, P^a.
\eea
Using
\begin{align}
&dg g^{-1}=dv v^{-1} + dx_a P^a-[dv v^{-1},x_a P^a] &
&\text{and} &
&du
u^{-1}
 = (1-\Ad(u))dv v^{-1},
\end{align}
one computes
\bea
\Omega= \langle du u^{-1}\wedge dx_a P^a\rangle-\langle du u^{-1}\wedge du
u^{-1},x_a P^a \rangle -s\langle v^{-1}dv\wedge v^{-1}dv, P_0\rangle,
\eea
which can be written as the exterior
derivative of the symplectic potential
\bea
\theta=-\langle du u^{-1},x_aP^a\rangle -s \langle dv v^{-1} ,
\hat p_a P^a\rangle.
\eea
This agrees with the symplectic potential derived  in
\cite{MW} in the spin-less case, provided we identify  their
$u$ with our $u^{-1}$, and generalises this expression
to the case of non vanishing spin. Finally, note that the symplectic
potential can be written very compactly as
\bea
\theta=-\langle dv v^{-1} ,
j_a P^a\rangle.
\eea

\section{The phase space of N massive particles
on a genus g  surface with boundary}
\label{phasespace}

\subsection{Boundary conditions and invariance transformations}

After describing the phase space of an open universe with a
single particle, we extend our description to a general spacetime
manifold  $M\approx\mathbb{R}\times S_{g,n}^\infty$ with
 $g$ handles and $n$ particles of
masses $\mu_i\in(0,2\pi)$ and spins $s_i$.
As we did for the single particle universe,
we must first impose an appropriate boundary
condition on the gauge fields at spatial infinity
and require the corresponding asymptotic behaviour
of the gauge transformations. Again, our choice of the
boundary condition is modelled after the
boundary condition in the metric formalism.
In a  reference frame where the centre of mass of  the universe is at
rest at the origin,
the physically sensible boundary condition at
spatial infinity is the requirement that the metric be asymptotically
conical of the form \eqref{conem} \cite{Matschull1}, i.~e.~that the
universe asymptotically appears like a single particle. Because of
the invariance of the conical metric \eqref{conem} under rotations
and time translations, the centre of mass condition does not uniquely
determine a reference frame.  However, we can imagine using a distant
particle to fix the orientation and time origin of a
distinguished centre of mass frame. This is analogous to the use
of distant fixed stars for selecting  a reference frame in  our (3+1)
dimensional universe.

As explained in the single particle
 discussion in Sect.~\ref{oneparticle},
we do not need to
restrict attention to centre of mass frames
in the Chern-Simons formulation. Rather we impose the
boundary condition that the gauge field at spatial infinity be
related to a connection of the form \eqref{gaugefield} 
by  a gauge transformation given
 asymptotically by a
constant element of the
Poincar\'e group, which physically represents
Poincar\'e transformations of the observer relative
to the  distinguished   centre of mass frame.
In a  universe with several particles and possibly handles, the parameters
$\mu$ and $s$ in \eqref{dreib} stand for the total mass and spin
of the universe.
 Note that we do not impose a fixed value for
$\mu$ and $s$ at this stage. 
However, we will find in Sect.~5.2 that
$\mu$ and $s$ remain constant during the time evolution of the
 universe. Our boundary condition can be rephrased by saying that 
 the  holonomy around the boundary representing
infinity is  in an elliptic conjugacy  class.
The equations of motion guarantee that  the class remains 
fixed during the time evolution.
Our strategy in  the following discussion
of the phase space will be to work in a centre of mass frame and
then generalise our results  to a general frame.

The starting point for our definition of the phase space of
$\PPoi$-Chern-Simons theory is the space  of solutions
of the equations of motion that satisfy the boundary condition. 
In a  centre of mass frame this is the space $\cala_\infty$
 of  flat connections on
$S^\infty_{g,n}$ which are of the form
 $A_\infty$
 \eqref{gaugefield}  for some value of $\mu$ and $s$ 
in an open neighbourhood of  the boundary. The physical phase space is
obtained as a quotient of this space by identifying solutions which
represent the same physical state. This means that we have to determine for
each invariance of the theory, i.~e.~for each bijection  of
the space  $\cala_\infty$ to itself,
 if it has
an interpretation as a gauge transformation  to be divided
out of the phase space or gives rise to a
physically meaningful symmetry transformation between different states.

The first type of invariance transformation in $\PPoi$-Chern-Simons
theory are gauge
transformations \eqref{gaugetransf} which are  compatible with the
boundary conditions at spatial infinity. 
In a centre of mass frame they are given by  the group  $\calp_\infty$
 of $\PPoi$-valued functions on $S^\infty_{g,n}$
which are a  constant
 rotation or time-translation in a
neighbourhood of  the boundary and act on the gauge
 field according to \eqref{gaugetransf}. 
It follows from the fact that  $\PPoi$ is contractible
(and in particular, simply connected) that  all of these transformations
are small,  i.~e.~connected to the identity transformation and obtained
by exponentiating infinitesimal transformations. Asymptotically
trivial transformations which are the identity in a neighbourhood of
the boundary form a subgroup $\calp\subset\calp_\infty$.
They are  generated
 by a gauge constraint and transform different descriptions of the
 same physical state into each other. As redundant
 transformations without physical meaning, they have to be divided out
 of the phase space, and field configurations related by them have to
 be identified. The situation is different for asymptotically
 nontrivial Chern-Simons gauge transformations. They are
 not generated directly by a gauge constraint but would require an
 additional boundary term in the action \cite{regte}, \cite{giu1}.
 They do not correspond to gauge degrees of freedom but give 
rise to symmetries acting on the phase
 space and have a physical interpretation of Poincar\'e
 transformations with respect to the distinguished centre of mass
frame.

The second type of invariance present
in $\PPoi$-Chern-Simons theory
arises from the fact that it is a topological theory. It is
invariant
under the group of all orientation preserving
diffeomorphisms compatible with the boundary conditions at spatial
infinity. This is the group  $\cald_\infty$ of orientation preserving
diffeomorphisms that reduce to a global spatial rotation in an
open neighbourhood of the boundary. Its elements act on the space
$\cala_\infty$ via pull back with the inverse
\bea
\label{diffact}
A\in \cala_\infty\rightarrow(\phi^{-1})^*A
=:\phi_*A\qquad\qquad\phi\in\cald_\infty\,.
\eea
Among the elements of $\cald_\infty$ we further distinguish the subgroup
$\cald\subset\cald_\infty$ of asymptotically 
trivial diffeomorphisms, i.~e.~diffeomorphisms of
 $S_{g,n}^\infty$ which keep
the  boundary  fixed pointwise. Diffeomorphisms in $\cald$ which are
connected to the identity are called small diffeomorphisms and form a
normal subgroup $\cald\subset\cald_0$.
In a Chern-Simons theory, diffeomorphisms are not a priori related to
gauge transformations and there is no reason to consider them as
redundant transformations between field configurations describing the
same physical state. However, Witten showed in \cite{Witten1} that
infinitesimal diffeomorphisms can on-shell be written as infinitesimal
Chern-Simons gauge transformations as follows.  The infinitesimal
transformation of a flat connection $A$ generated by a vector field
$V$ is given by the Lie derivative
\bea
\label{infdiff}
\delta A={\mathcal L}_V A\,.
\eea
Using the formula  ${\mathcal L}_V=d i_V+ i_V d$ and the 
flatness of $A$, this can be written as an
infinitesimal gauge transformation
\bea
\label{equivalence}
{\mathcal L}_V A=d\lambda + [A,\lambda]
\eea
with generator
\bea
\label{generators}
\lambda =i_V A,
\eea
where $i_V$ denotes the contraction with $V$.
By exponentiating, we see that any two field configurations $A$ and
$A'$ related by a small, asymptotically trivial diffeomorphism are also
related by an asymptotically trivial Chern-Simons gauge transformation
and therefore have to be identified. Small
diffeomorphisms that reduce to global rotations at the boundary
correspond to asymptotically nontrivial Chern-Simons gauge transformations
and therefore represent  physical transformations with respect to the
centre of mass frame. However, the correspondence between
diffeomorphisms and Chern-Simons gauge transformations does not hold for
large diffeomorphisms, i.~e.~diffeomorphisms that are not
infinitesimally generated. Those diffeomorphisms are not related to
gauge degrees of freedom but are 
physically meaningful transformations acting
on the phase space.

We conclude that the physical phase space of $\PPoi$-Chern-Simons 
theory is obtained from the space
$\cala_\infty$ of flat gauge connections that satisfy the boundary
conditions by identifying connections $A$, $A'\in\cala_\infty$ if and 
only if they are related by an
asymptotically trivial Chern-Simons gauge transformation. This implies
division by small, asymptotically trivial diffeomorphisms, whereas
large diffeomorphisms and asymptotically nontrivial Chern-Simons
transformations give rise to physical symmetries. We will return to
this question and give a more mathematical treatment in Sect.~5.1,
where we determine the symmetry group of the theory.

 A final general comment we should make concerns the
 dreibein $e_a$ in the decomposition of a
connection $A$ according to \eqref{Cartan}.
In Einstein's metric formulation of gravity
the dreibein is assumed to be invertible,
but there is no such requirement in Chern-Simons
theory. In fact, as pointed out and discussed in \cite{Matschull2},
gauge orbits under  $\PPoi$ gauge transformations may
pass through a connection with a degenerate dreibein.
In \cite{Matschull2} it is argued that this leads to the
 identification of spacetimes in the Chern-Simons formulation of (2+1) gravity
which would  be regarded as physically distinct in the metric formulation.
We will address this issue in the final section of this paper,
after we have given a detailed description of the phase space
of $\PPoi$-Chern-Simons theory.

\subsection{The description of the space time by a graph}
 The description of the phase space in the 
formalism of  Fock and Rosly depends on a  graph.
As in the case of a single massive particle, the
graphs used in the construction of 
the Poisson structure on the moduli space have a
 intuitive physical interpretation. For a (sufficiently refined)
graph that is embedded into the surface $S_{g,n}^\infty$ in such a way
that the punctures representing massive particles lie on different faces
of the graph and at least one vertex is mapped to the boundary at
spatial infinity, the  vertices of the graph can be thought of as
observers located at different spacetime points. As the spacetime
manifold is locally flat,
 the reference frames of these observers should be
inertial frames related by Poincar\'e transformations. These
Poincar\'e transformations are given by the graph connection. The
$\PPoi$-element $g=(v,{\bf y})$ assigned to an (oriented) edge can be
interpreted as the Poincar\'e transformation relating the reference
frames of the observers at its
ends: the translation vector ${\bf y}$ describes the shift
in position and time coordinate, the Lorentz transformation
$v$ the relative orientation of their coordinate axes and their
relative velocity. The observers located at different points in
the spacetime can determine topology and matter content of their
universe by exchanging information about these Poincar\'e
transformations relating their reference frames.  Due to the
flatness of the graph connection, observers situated at vertices
surrounding a face without particles will
find that the (ordered) product of the Poincar\'e transformations
along the boundary of the face is equal to the identity. If the face
contains a massive particle, the same procedure yields an element
of the $\PPoi$-conjugacy class determined by mass and spin of the
particle, thus allowing a measurement of these quantities.

The interpretation of graph connections as Poincar\'e transformations
between reference frames of observers situated at the vertices
is complemented by  the physical interpretation of graph gauge
transformations \eqref{graphgauge}. A graph gauge transformation
 involves an individual $\PPoi$ transformation at each vertex.
As the Poincar\'e transformations relating adjacent observers have
the right transformation property \eqref{graphgauge}, the graph gauge
transformations can be interpreted as changes of the reference
frame for each observer. In this interpretation it is obvious that all
transformations not affecting the vertices at the boundary correspond
to gauge degrees of freedom. They do not alter the physical state of
the universe but only its description in terms of reference frames of
different observers. The situation is different, however, for a
vertex
situated at the boundary.
With the interpretation of such a vertex as an observer, the
 graph gauge transformations affecting the vertex
are seen to (uniquely) characterise the asymptotic
symmetries discussed above. They correspond to
rotations, boosts and translations of the observer
 with respect to the distinguished centre of mass frame
of the universe.

\subsection{The phase space of $n$ massive particles and $g$ handles}

Taking into account the special role of the boundary,
the most efficient Fock-Rosly graph for a surface
$S_{g,n}^\infty$ is a set of curves representing  the generators
of the fundamental group with basepoint $z_\infty$ on the boundary as
pictured in Fig.~3.

\vbox{
\vskip .3in
\input epsf
\epsfxsize=12truecm
\centerline{
\epsfbox{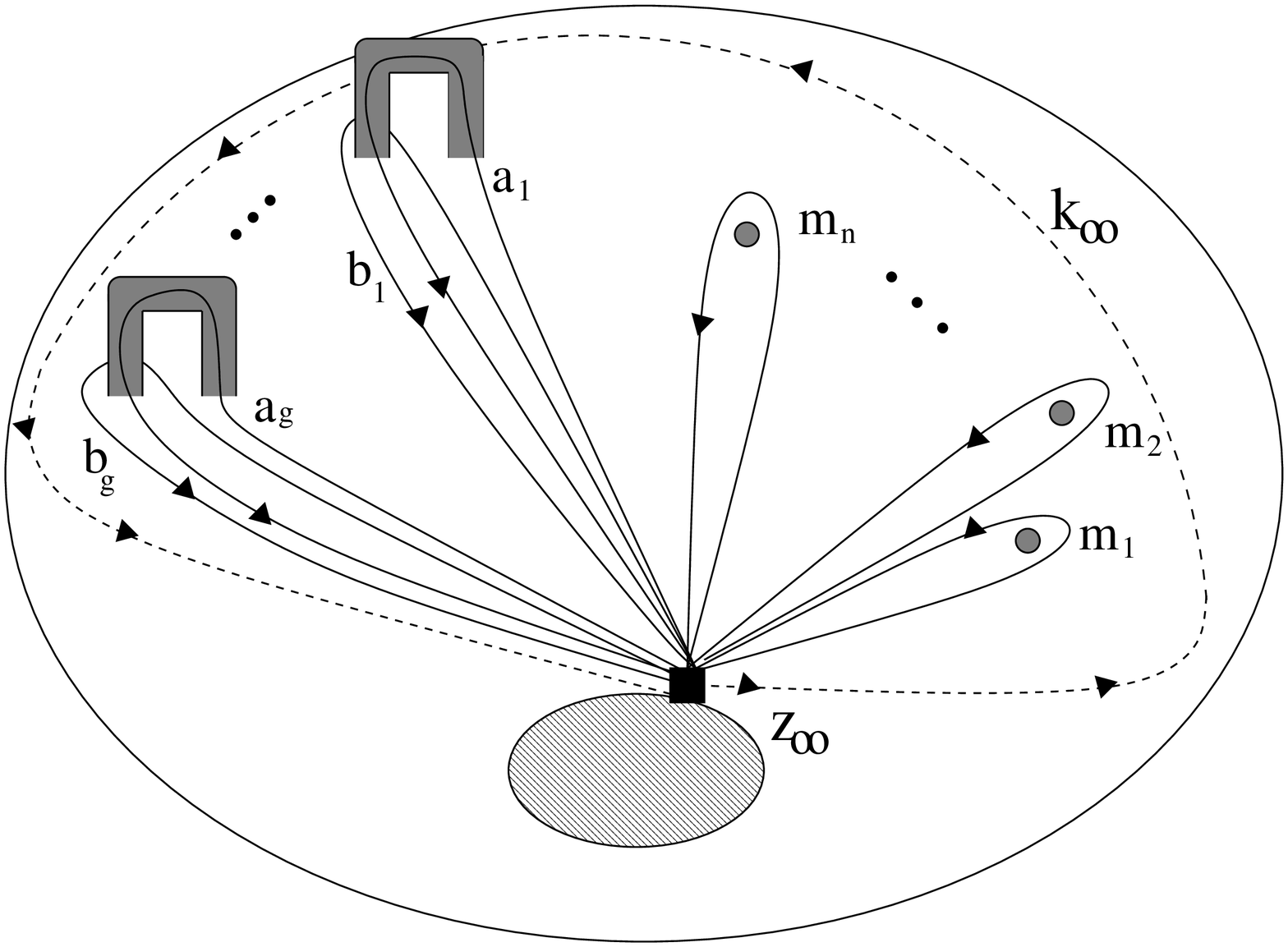}
}
\bigskip
{
\centerline{\bf Fig.~3 }
}
\centerline{\footnotesize The generators of the
fundamental group of the surface $S_{g,n}^\infty$}
\bigskip}

 The fundamental group is generated by the equivalence classes of the
 curves $m_1,\ldots,m_n$ around each particle, curves
 $a_1,b_1,\ldots,a_g,b_g$ around each handle and a curve around the
 boundary at spatial infinity, subject to a relation similar to
 \eqref{relation}. By solving the relation, the fundamental group
 can be presented as
 the free group generated by the curves around the particles and handles
\bea
\pi_1(z_\infty,S_{g,n}^\infty)=<m_1,\ldots,m_n,a_1,b_1,\ldots,a_g,b_g>.
\eea
 The holonomy
along these curves assigns elements $M_i\in \PPoi$, $i=1,\ldots,n$, $A_j$,
$B_j\in \PPoi$, $j=1,\ldots,g$ of the gauge group $\PPoi$ to each of
the generators $m_i$, $a_j$, $b_j$. Asymptotically nontrivial
Chern-Simons gauge transformations
\eqref{gaugetransf} induce graph gauge transformations that act
on these group elements
via simultaneous conjugation. As explained in the previous
subsection, these are  physical symmetries
and  not divided out of the physical phase space.

The same is
true for the group  of asymptotically trivial, large diffeomorphisms.
Mathematically,  this is called the
mapping class group and defined as
\bea
\label{mcg}
 \text{Map}(S_{g,n}^\infty)= \mathcal{D}/\mathcal{D}_0.
\eea
In our description of the spacetime by means of an embedded graph,
asymptotically trivial, large diffeomorphisms map
the graph to a different, topologically inequivalent graph.
If we include the large diffeomorphisms as symmetries acting on the
phase space, our phase space is not simply given as the quotient of
the space of graph connections modulo (asymptotically trivial)
graph gauge transformations
corresponding to a {\em fixed graph}. Rather, we consider
topologically
distinct graphs and include  an additional
label specifying the graph. In the case where the graph is
given by a set of generators of
 the fundamental group, this can be made more explicit.

The starting point for our definition of the phase  space
is one fixed graph which consists of the generators of the
fundamental group $\pi_1(z_\infty,S_{g,n}^\infty)$  shown in
Fig.~3.
With respect to this graph, the phase space is simply the product
\bea
\label{startingpoint}
\mathcal{C}_1\times \cdots\times  \mathcal{C}_n
\times (\PPoi)^{2g} \;,
\eea
 where
\bea
\label{conjug}
\mathcal{C}_i=\{g\,(\exp(-\mu_iJ^0),-(s_i,0,0)^t)\,g^{-1}\;|\;
g\in \PPoi\}\qquad i=1,\ldots,n
\eea
are the $\PPoi$-conjugacy classes determined by masses and spins of
the particles.

The mapping class group $ \text{Map}(S_{g,n}^\infty)$ acts on
$\pi_1(z_\infty,S_{g,n}^\infty)$
and induces a subgroup of the  automorphism group of
$\pi_1(z_\infty,S_{g,n}^\infty)$ \cite{birmgreen},\cite{birmorange}.
The  graphs related to the one depicted in
Fig.~3  by an element of $ \text{Map}(S_{g,n}^\infty)$
are therefore conveniently labelled by that element. Finally,
by associating to every element of the mapping class group
the induced permutation of the punctures $z_1,\ldots z_n$,
we get a homomorphism into the permutation group $P_n$
\bea
\label{permute}
\pi: \text{Map}(S_{g,n}^\infty)\rightarrow P_n,
\eea
in terms of which the phase space is defined as
\bea
\label{npartphspace}
\mathcal{M}_{\pi_1}&=&\{
\left(\sigma,(M_1,\ldots,M_n,A_1,B_1,\ldots,A_g,B_g)
\right)
\in \text{Map}(S_{g,n}^\infty)
\times (\PPoi)^{n+2g}
 \nonumber \\
&&|\;M_i\in\mathcal{C}_{\pi(\sigma)(i)},\;
i=1,\ldots,n\}.
\eea

We use the following parametrisation of the phase space.
As the holonomies of the loops around the punctures are
elements of fixed, elliptic $\Poi$-conjugacy classes and the
particle's masses are in the interval $(0,2\pi)$, they can be
parametrised by
\begin{align}
&M_i=(u_\mi,\ba^\mi)\qquad\qquad\text{with}&
&u_\mi=\exp(-p^a_\mi J_a),\;\ba^\mi=-\Ad(u_\mi)\bj^\mi,\\
& & &p_\mi^0>0,\,\bp_\mi^2=\mu_i^2,\;\mu_i\in(0,2\pi).\nonumber
\end{align}
In terms of the parameters  $p^\mi_a$, $j^\mi_a$ the constraints then
take the form \eqref{massspin}
\bea
\label{constraint2}
p_a^\mi p^a_\mi\approx \mu_i^2\qquad p_a^\mi j^a_\mi\approx \mu_i s_i
 \qquad i=1,\ldots,n,
\eea
in agreement with the interpretation of $\mu_i\in (0,2\pi)$ and $s_i$ as
masses and spins of the particles.
There are no such constraints for the holonomies of the loops around
the handles. They can be elliptic, parabolic or hyperbolic elements of
$\PPoi$ and therefore have to be parametrised in the more general form
\eqref{lparam}
\begin{align}
&A_i=(u_\ai,\ba_\ai) & &\text{with}\;u_\ai=\exp(-2\pi n_\ai J_0)
\exp(-p^a_\ai J_a),\;\ba_\ai=-\Ad(u_\ai)\bj^\ai\\
&B_i=(u_\bi,\ba_\bi) & &\text{with}\;u_\bi=\exp(-2\pi n_\bi J_0)
\exp(-p^a_\bi J_a),\;\ba_\bi=-\Ad(u_\bi)\bj^\bi,\nonumber
\end{align}
where $n_\ai,n_\bi\in \mathbb{Z}$.

\subsection{Poisson structure}
 As the graph given by the fundamental group has only one vertex, the
 Poisson structure on the phase space involves the choice of a single
$r$-matrix. It is given by the bivector introduced in \cite{AS}
\begin{align}
\label{fundbivect}
B_{FR}=&\sum_{i=1}^n r^{\alpha\beta}\left(\frac{1}{2}R_\alpha^{M_i}
\wedge R_\beta^{M_i}+\frac{1}{2}L_\alpha^{M_i}\wedge L_\beta^{M_i}+
R_\alpha^{M_i}\wedge L_\beta^{M_i}\right)\\
+&\sum_{i=1}^gr^{\alpha\beta}\bigg(\frac{1}{2}\left( R_\alpha^{A_i}
\wedge R_\beta^{A_i}+L_\alpha^{A_i}\wedge L_\beta^{A_i}+R_\alpha^{B_i}
\wedge R_\beta^{B_i}+L_\alpha^{B_i}\wedge L_\beta^{B_i}\right)\nonumber\\
&\qquad\qquad+R_\alpha^{A_i}\wedge (R_\beta^{B_i}+L_\beta^{A_i}+
L_\beta^{B_i})+R_\alpha^{B_i}\wedge (L_\beta^{A_i}+L_\beta^{B_i})+
L_\alpha^{A_i}\wedge L_\beta^{B_i}\bigg)\nonumber\\
+&\sum_{1\leq i<j\leq n}r^{\alpha\beta} (R_\alpha^{M_i}+
L_\alpha^{M_i})\wedge(R_\beta^{M_j}+L_\beta^{M_j})\nonumber\\
+&\sum_{1\leq i<j\leq g}r^{\alpha\beta}(R_\alpha^{A_i}+
L_\alpha^{A_i}+R_\alpha^{B_i}+L_\alpha^{B_i})
\wedge(R_\beta^{A_j}+L_\beta^{A_j}+R_\beta^{B_j}+L_\beta^{B_j})\nonumber\\
+&\sum_{i=1}^n\sum_{j=1}^g
r^{\alpha\beta}(R_\alpha^{M_i}+L_\alpha^{M_i})
\wedge(R_\beta^{A_j}+L_\beta^{A_j}+R_\beta^{B_j}+L_\beta^{B_j})\nonumber\,,
\end{align}
where $M_1,\ldots,M_n$ denote the holonomies  around the particles,
$A_1,B_1,\ldots,A_g,B_g$  the holonomies
corresponding to the handles and the cilium at the
vertex $z_\infty$ is chosen such that the order of the incident edges is
\begin{align}
m_1<\ldots<m_n<a_1<b_1<a_1^{-1}<b_1^{-1}<
\ldots<a_g<b_g<a_g^{-1}<b_g^{-1}.\nonumber
\end{align}
As there is no division by global conjugation, the correct
$r$-matrix  has again to be determined from a {\em physical}
  requirement,
namely, that the Poisson brackets of total momentum $\bp_{tot}$ and
total angular momentum $\bj_{tot}$ of the universe are
those of a free particle and act as the infinitesimal generators of
asymptotic Poincar\'e transformations.
As in the case of a single particle, this requires the $r$-matrix
$r=P_a\otimes J^a$. Taking into account that the elements $\exp(-2\pi
n J_0)$ with $n\in\mathbb{Z}$ are central in
$\PPoi$, we obtain
expressions of the form \eqref{vecfields} for the right and left
invariant vector fields corresponding to the generators of the
fundamental group.
Inserting these fields and the $r$-matrix
 into equation \eqref{fundbivect}, we derive the Poisson structure:

\begin{align}
\label{singlebr}
 &\{p_{M_i}^a,p_{M_i}^b\}=0 & &i=1,\ldots,n\\
 &\{j_{M_i}^a,p_{M_i}^b\}=-\epsilon^{abc}p_c^{M_i}\nonumber\\
 &\{j_{M_i}^a,j_{M_i}^b\}=-\epsilon^{abc}j_c^{M_i}\nonumber\\
\nonumber\\
\label{mixbr}
&\{j_{X}^a,p_{Y}^b\}=
-(1-\text{Ad}(u_{X}))_d^{\;\;a}\epsilon^{bcd}p_c^{Y} &
&X<Y,\\
&\{j_{X}^a,j_{Y}^b\}=
-(1-\text{Ad}(u_{X}))_d^{\;\;a}\epsilon^{bcd}j_c^{Y} & &X,Y\in\{M_1,\ldots,M_n,A_1,\ldots,B_g\}\nonumber\\
&\{p_{X}^a,j_{Y}^b\}=0\nonumber\\
&\{p_{X}^a,p_{Y}^b\}=0\nonumber\\
\nonumber\\
\label{handlebr}
 &\{p_{A_i}^a,p_{A_i}^b\}=0 & &i=1,\ldots,g\\
 &\{j_{A_i}^a,p_{A_i}^b\}=-\epsilon^{abc}p_c^{A_i}\nonumber\\
 &\{j_{A_i}^a,j_{A_i}^b\}=-\epsilon^{abc}j_c^{A_i}\nonumber\\
 &\{p_{B_i}^a,p_{B_i}^b\}=0 \nonumber \\
 &\{j_{B_i}^a,p_{B_i}^b\}=-\epsilon^{abc}p_c^{B_i}\nonumber\\
 &\{j_{B_i}^a,j_{B_i}^b\}=-\epsilon^{abc}j_c^{B_i}\nonumber\\
&\{j_{A_i}^a,j_{B_i}^b\}=-\epsilon^{abc}j_c^{B_i}\nonumber\\
&\{p_{A_i},p_{B_i}\}=0\nonumber\\
&\{j_{A_i}^a,p_{B_i}^b\}=-\left(1+\Ad({u_\ai^{-1}})
\cdot\frac{\text{Ad}(u_{B_i})}{1-\text{Ad}(u_{B_i})}\right)^a_{\;\;d}
\epsilon^{bcd}p_c^{B_i}= & &-\epsilon^{abc}p_c^\bi+\Ad(u^{-1}_\ai)^a_{\;\;c}
(T^{-1}(\bp_\bi))^{cb}\nonumber\\
&\{j_{B_i}^a,p_{A_i}^b\}=\left(\frac{\text{Ad}(u_{A_i})}
{1-\text{Ad}(u_{A_i})}\right)^a_{\;\;d}\epsilon^{bcd}p_c^{A_i}=
-(T^{-1}(\bp_\ai))^{ab},\nonumber
\end{align}
where $M_1<\cdots<M_n<A_1,B_1<\cdots<A_g,B_g$.
The maps $T(\bp_\ai),T(\bp_\bi)$ are given by
\eqref{Ttrans} and their inverses by \eqref{tdef}.
Note that the Poisson structure is not the direct product of the
Poisson structures of different particles and handles but mixes their
contributions. Furthermore, it can be seen from equations
\eqref{singlebr} to \eqref{handlebr} that the mass and spin constraints
\eqref{constraint2} for the particles Poisson commute with all
functions on the phase space. This is not true for the corresponding
expressions for the handles, as we have from \eqref{handlebr}
\bea
\{\bp_\bi\cd\bj_\bi,\bp_\ai^2\}=-\{\bp_\ai\cd\bj_\ai,\bp_\bi^2\}&=&
2\bp_\ai\cd\bp_\bi.
\eea
In fact, we will see at the end of this section that the
Poisson
structure corresponding to a single handle has no Casimir functions
and is symplectic.

\subsection{The decoupling transformation}

In \cite{AMII},
 Alekseev and Malkin discovered a bijective transformation of the
 moduli
space of flat $G$ connections  that maps its symplectic structure
 onto the direct product of $n$ copies of the symplectic form
 on the conjugacy
classes of $G$ and $g$ copies of the symplectic  form on the Heisenberg
double $\mathcal{D}_+(G)$ as defined in the appendix.
They proved their result for the case of a compact
 semi-simple Lie group $G$ and an underlying closed  surface with punctures.
For our purposes we need an extension of their result to non-compact
gauge groups and a  surfaces with boundary. In order
to state  our result as simply as possible we define the inverse of
the transformation in \cite{AMII}.
With the  factorisation
\eqref{factoring} $(u,\ba)=(u,\ba)_+(u,\ba)_-^{-1}$
discussed in the appendix,
it is given by
\begin{align}
K^{-1}:\;&\mathcal{C}_{1}\times\cdots\times \mathcal{C}_n\times
\underbrace{\mathcal{D}_+(\PPoi)\times\cdots\times\mathcal{D}_+
(\PPoi)}_{g\times}\longrightarrow \mathcal{C}_{1}\times\cdots\times
\mathcal{C}_n\times\underbrace{\PPoi\times\cdots\times\PPoi}_{2g\times}
\nonumber
\end{align}
\begin{align}
\label{septrafo}
\qquad &\mip=(u_\mip,\ba_\mip)& &\longrightarrow & \mi&=(u_\mi,\ba_\mi)\\
&\quad & &\quad &  &=M'_{(n+2g)-}\cdots
M'_{(i+1)-}\cdot\mip\cdot M'^{-1}_{(i+1)-}\cdots M'^{-1}_{(n+2g)-}\nonumber\\
&\aip=(u_\aip,\ba_\aip) & &\longrightarrow & \ai&=(u_\ai,\ba_\ai)\nonumber\\
&\quad & &\qquad &  &=M'_{(n+2g)-}\dots
M'_{(n+2i+1)-}\cdot\aip\cdot M'^{-1}_{(n+2i)-}\cdots
M'^{-1}_{(n+2g)-}\nonumber\\
&\bip=(u_\bip,\ba_\bip) & &\longrightarrow & \bi&=(u_\bi,\ba_\bi)\nonumber\\
&\quad & &\qquad &  &=M'_{(n+2g)-}\cdots
M'_{(n+2i+1)-}\cdot\bip\cdot M'^{-1}_{(n+2i)-}\cdots
M'^{-1}_{(n+2g)-}\nonumber
\end{align}
with
\bea
M'_{(n+2i)}:=\bip\aip^{-1}\qquad \text{and} \qquad M'_{(n+2i-1)}:=
\bip^{-1}\aip\;.\nonumber
\eea
Then our result can be stated as follows.
\begin{theorem}
\begin{enumerate}
\item The transformation $K^{-1}$ given by \eqref{septrafo} transforms
the momentum and
    angular momentum three-vectors of particles and handles according
    to
{\em
\begin{align}
\label{pjseptrafo}
&\bp^\mi=\bp^\mip,\qquad\bp^\ai=\bp^\aip,\qquad\bp^\bi=\bp^\bip\\
&u_\mi=u_\mip,\qquad u_\ai=u_\aip,\qquad u_\bi=u_\bip\nonumber\\
&\bj^\mi=\bj^\mip+(1-\Ad(u_\mip^{-1}))\left(\sum_{j=i+1}^n
  \bj^\mjp\right)
\nonumber\\
&\qquad+(1-\Ad(u_\mip^{-1}))\left(\sum_{j=1}^g
\left(1-\Ad({u_\ajp})\right)\bj^\ajp+\left(\Ad({u_\ajp})-\Ad({u^{-1}_\ajp
  u_{\bjp}})\right)\bj^\bjp\right)\nonumber\\
&\bj^\ai=\bj^\aip-\Ad({u_\aip})(\bj^\aip-\bj^\bip)\nonumber\\
&\qquad+(1-\Ad({u_\aip^{-1}}))\left(\sum_{j=i+1}^g
\left(1-\Ad({u_\ajp})\right)\bj^\ajp+\left(\Ad({u_\ajp})-\Ad(u^{-1}_\ajp
  u_\bjp)\right)\bj^\bjp\right)\nonumber\\
&\bj^\bi=\bj^\bip-\Ad({u_\aip})(\bj^\aip-\bj^\bip)\nonumber\\
&\qquad+(1-\Ad({u_\bip^{-1}}))\left(\sum_{j=i+1}^g
\left(1-\Ad({u_\ajp})\right)\bj^\ajp+\left(\Ad({u_\ajp})-\Ad({u^{-1}_\ajp
  u_\bjp})\right)\bj^\bjp\right).\nonumber
\end{align}}
\item Its inverse $K$ maps the Poisson structure given by
  \eqref{singlebr} to \eqref{handlebr} to the direct product of $n$
symplectic forms on the conjugacy
classes and $g$ Poisson structures on the Heisenberg double $\mathcal{D}_+(g)$

{\em \begin{align}
\label{sepbrpart}
&\{p_\mip^a,p_\mip^b\}=0\qquad\qquad\qquad\qquad\qquad i=1,\ldots,n\\
&\{j_\mip^a,p_\mip^b\}=-\epsilon^{abc}p_c^\mip \nonumber\\
&\{j_\mip^a,j_\mip^b\}=-\epsilon^{abc}j_c^\mip\nonumber\\
\nonumber\\
\label{sepbrmix}
&\{j_{X}^a,p_{Y}^b\}=0
\qquad\qquad X<Y,\; X,Y\in\{M'_1,\ldots,M'_n,A'_1,\ldots,B'_g\}\\
&\{j_{X}^a,j_{Y}^b\}=0\nonumber\\
&\{p_{X}^a,j_{Y}^b\}=0\nonumber\\
&\{p_{X}^a,p_{Y}^b\}=0\nonumber\\
\nonumber\\
\label{sepbrhandle}
&\{p_{\aip}^a,p_{\aip}^b\}=0
\qquad\qquad\qquad\qquad\qquad\qquad\qquad i=1,\ldots,g\\
&\{j_{\aip}^a,p_{\aip}^b\}=-\left(\frac{1}{1-\Ad({u_{\aip}})}
\right)^a_{\;\;d}\epsilon^{bcd}p_c^{\aip}
=\Ad(u_{\aip}^{-1})^a_{\;\;c}(T^{-1}(\bp_{\aip}))^{cb}\nonumber\\
&\{j_{\aip}^a,j_{\aip}^b\}=-\epsilon_{abc}j_c^{\aip}\nonumber\\
&\{p_{\bip}^a,p_{\bip}^b\}=0\nonumber\\
&\{j_{\bip}^a,p_{\bip}^b\}=-\left(\frac{1}{1-\Ad({u_{\bip}})}
\right)^a_{\;\;d}\epsilon^{bcd}p_c^{\bip}
=\Ad(u_{\bip}^{-1})^a_{\;\;c}(T^{-1}(\bp_{\bip}))^{cb}\nonumber\\
&\{j_{\bip}^a,j_{\bip}^b\}=-\epsilon_{abc}j_c^{\bip}\nonumber\\
&\{p_{\aip}^a,p_{\bip}^b\}=0\nonumber\\
&\{j_{\aip}^a,p_{\bip}^b\}=-
\left(\left(1+\Ad({u^{-1}_{\aip} u_{\bip}})\right)\cdot\frac{1}
{1-\Ad(u_{\bip})}\right)^a_{\;\;d}\epsilon^{bcd}p_c^{\bip}\nonumber\\
&\qquad\qquad=\left(\Ad(u^{-1}_{\bip})+\Ad(u^{-1}_{\aip})\right)^a_{\;\;c}
(T^{-1}(\bp_{\bip}))^{cb}\nonumber\\
&\{j_{\aip}^a,j_{\bip}^b\}=-\epsilon^{abc}j_c^{\bip}\nonumber\\
&\{j_{\bip}^a,p_{\aip}^b\}=0\nonumber
\end{align}}
where, again, $M'_1<\cdots< M'_n<A'_1,B'_1<\cdots<A'_g,B'_g$, and the maps $T(\bp_\aip),T(\bp_\bip)$
are given by \eqref{Ttrans}, their inverses by \eqref{tdef}.
\end{enumerate}
\end{theorem}

{\bf Proof:} The result follows by a direct but lengthy computation.
The following steps and formulae are essential.
\begin{enumerate}
\item We obtain the transformation of the parameters $\bp$, $\bj$
  by inserting $(u,\ba)_-=(1,-\Ad(u^{-1})\ba)=(1,\bj)$ in
\eqref{septrafo}, which yields \eqref{pjseptrafo} by straightforward
  calculation. Further calculation proves that $K^{-1}$ is
indeed bijective and can be inverted.
\item   We show that the transformation $K^{-1}$ given by equation
  \eqref{pjseptrafo} transforms the brackets \eqref{sepbrpart} to
  \eqref{sepbrhandle} into the Poisson brackets \eqref{singlebr} to
  \eqref{handlebr}. Defining
\begin{align}
\label{jhandle}
&\bv^\aip=\bj^\aip-\Ad(u_\ai)(j^\aip-j^\bip)\\
&\bv^\bip=\bj^\bip-\Ad(u_\ai)(j^\aip-j^\bip)\nonumber\\
&\bj^{H'_i}=\left(1-\Ad(u_\ai)\right)\bj^\aip+
\left(\Ad(u_\ai)-\Ad(u_\ai^{-1}u_\bi)\right)\bj^\bip\nonumber
\end{align}
for $i=1,\ldots,g$, equations \eqref{pjseptrafo} read
\begin{align}
&\bj^\mi=\bj^\mip+(1-\Ad(u_\mip^{-1}))\left(\sum_{j=i+1}^n
  \bj^\mjp+\sum_{j=1}^g \bj^{H'_j}\right) & &i=1,\ldots,n\\
&\bj^\ai=\bv^\aip+(1-\Ad({u_\aip^{-1}}))\left(\sum_{j=i+1}^g
  \bj^{H'_j}\right) & &i=1,\ldots,g\nonumber\\
&\bj^\bi=\bv^\bip+(1-\Ad({u_\bip^{-1}}))\left(\sum_{j=i+1}^g
  \bj^{H'_j}\right) & &i=1,\ldots,g.\nonumber \end{align}
We insert these expressions into
  \eqref{sepbrpart} to \eqref{sepbrhandle} and
  calculate the resulting Poisson brackets, using
  the expression \eqref{adjoint} for the adjoint in terms of
  $\bp$, the $\Ad$-invariance of the epsilon-tensor as well as the
  following auxiliary formulas
\begin{align*}
&\{v^\aip_a,\Ad(u_\aip)_{bc}(v^\aip_c-v^\bip_c)\}=
\{v^\bip_a,\Ad(u_\aip)_{bc}(v^\aip_c-v^\bip_c)\}=
-\epsilon_{abc}\Ad(u_\aip)^{ce}(v^\aip_e-v^\bip_e)\\
&\{j^{H'_i}_a,j^{H'_i}_b\}=-\epsilon_{abc} j_{H'_i}^{c}\\
&\{j^{H'_i}_a,v^{X'}_b\}=-\epsilon_{abc}v_{X'}^c\\
&\{j^{H'_i}_a,p^{X'}_b\}=-\epsilon_{abc}p_{X'}^c\\
&\{\Ad(u_{X'})_{ab},j^{H'_i}_c\}=-\epsilon_{ace}
\Ad(u_{X'})_{\;\;b}^{e}-\epsilon_{bce}\Ad(u_{X'})_a^{\;\;e}
\qquad\qquad\text{for}
\;X'=A_i,B_i.
\end{align*}
After a rather complicated calculation, we obtain brackets
\eqref{singlebr} to \eqref{handlebr}. $\qquad\Box$
\end{enumerate}

We see  that the Poisson brackets \eqref{sepbrpart} are
just the direct sum of $n$ single particle brackets, i.~e.~of
$n$ copies of the Poisson structure on the $\PPoi$-conjugacy
classes. The Poisson brackets \eqref{sepbrmix} of three-momenta
and the angular momentum three-vectors associated to handles and
particles with those associated to different handles vanish.
The contribution \eqref{sepbrhandle} of each handle agrees with
the expression \eqref{hdbracket} for the symplectic structure
on the Heisenberg double given in the appendix.
We shall see in the next section that the total
angular momentum three-vector has a particularly simple expression in
 terms of the decoupled coordinates.

\section{Symmetries and the action of the mapping class group}
\label{symmetry}

\subsection{The symmetry group in the Chern-Simons formulation of
  (2+1) gravity}

 As explained in Sect.~4.1,
  two
flat connections in $\cala_\infty$
describe the same physical state if and only
if they can be transformed into each other by an asymptotically
  trivial Chern-Simons gauge transformation. Due to the relation
  \eqref{equivalence}, this is the case for connections related by
  small, asymptotically trivial diffeomorphisms, whereas
 large diffeomorphisms and asymptotically
nontrivial  $\PPoi$-
gauge transformations give rise to physical
symmetries. However, we have yet to determine exactly the symmetry
group of our model and its action on the physical phase space.

The
question of the symmetry group of gauge theories or general
relativity on a manifold with a boundary was  investigated by
Giulini  in \cite{giu1},\cite{giu2}. He classifies the transformations that map
the space of solutions to itself into three categories: the group
$\mathcal{T}^\infty$ of transformations compatible with the boundary
condition, the group $\mathcal{T}$ of transformations that are
trivial at the boundary and its identity component, the group
$\mathcal{T}_0$ of infinitesimally generated transformations that are
trivial at the boundary. Among these, only the transformations in
$\mathcal{T}_0$ are generated by the gauge constraints and
map solutions to physically equivalent solutions that should be
identified.
The other two types of transformations do not correspond to gauge
degrees of freedom but are related to physical symmetries. The quotient
$\mathcal{B}=\mathcal{T}^\infty/\mathcal{T}$ is the { asymptotic
 symmetry group}, reflecting the symmetries of the boundary
condition,
and the symmetry group $\mathcal{S}$ of the theory is given as
the quotient $\mathcal{S}=\mathcal{T}^\infty/\mathcal{T}_0$. Giulini
showed that these quotients are related by a bundle structure
\bea
\label{bundle}
\begin{array}{ccc}
\calt/\calt_0 & \xrightarrow{i} & \cals=\calt_\infty/\calt_0\\
& & \downarrow p\\
& & \calb=\calt_\infty/\calt.
\end{array}
\eea
If the bundle is trivial, the symmetry group is simply the product
$\cals=\calt/\calt_0\times\calb$ of the group $\calt/\calt_0$
and the asymptotic symmetry group. In general, this is not
the case and the symmetry group has a more complicated structure.

We now  apply the results of \cite{giu1} and \cite{giu2} to our model
of $\PPoi$-Chern-Simons theory on a surface $S^\infty_{g,n}$, starting
with the definition of the groups $\calt_\infty$, $\calt$
and $\calt_0$. The group $\calt_\infty$ of transformations compatible with the
boundary conditions combines  the group $\cald_\infty$ of
diffeomorphisms that reduce to global rotations and
the group $\calp_\infty$ of $\PPoi$-transformations which reduce to
constant spatial rotations and time-translations at the boundary.
 The actions of these groups on the space $\cala_\infty$ are given by
equation \eqref{diffact} and \eqref{gaugetransf},
respectively. Considering their semidirect product
$\cald_\infty \ltimes {\calp_\infty}$
with group multiplication
\bea
(\phi_1,\gamma_1)(\phi_2,\gamma_2)=
(\phi_1\circ\phi_2,\gamma_1\left(\gamma_2\circ\phi_1^{-1}\right))
\eea
and combining their actions acoording to
\bea
A \in\cala_\infty \rightarrow \gamma\phi_*A\gamma^{-1} + 
\gamma d\gamma^{-1}\qquad\text{for}\;\phi\in\cald_\infty,\,
\gamma\in\calp_\infty,
\eea
we obtain a homomorphism
\bea
\label{rhoinf}
\rho_\infty: \cald_\infty \ltimes {\calp_\infty} \rightarrow 
\text{Aut}(\cala_\infty)
\eea
into the space Aut($\cala_\infty$) of bijections of $\cala_\infty$ to itself.
The group $\calt_\infty$ is obtained from the semidirect product
 $\cald_\infty\ltimes\calp_\infty$ by identifying diffeomorphisms and
Chern-Simons gauge transformations that are identical {\em as maps} of
$\cala_\infty$ to itself, i.~e.~by dividing out the kernel of $\rho_\infty$
\bea
\label{inftrans}
\calt_\infty = (\cald_\infty \ltimes {\calp_\infty}) /\mbox{ker}\;
\rho_\infty\,.
\eea
 Note in this context that the equivalence
\eqref{equivalence} between infinitesimal diffeomorphisms and
Chern-Simons gauge transformations only holds for a given connection
$A\in\cala_\infty$ and does not imply that these transformations
induce the same maps of $\cala_\infty$ via the homomorphism $\rho_\infty$.

The group $\calt$ of asymptotically trivial  invariance transformations
combines   asymptotically trivial diffeomorphisms  with 
asymptotically trivial Chern-Simons gauge transformations. We define 
the  homomorphism
$\rho: \cald \ltimes \calp \rightarrow \text{Aut}(\cala_\infty)$
as the restriction of $\rho_\infty$  \eqref{rhoinf} to $\cald\ltimes \calp$.
Then the   group $\calt$ is given by
\bea
\label{trans}
\calt = (\cald \ltimes \calp) /\mbox{ker}\;\rho\,.
\eea

Finally,
the group $\calt_0$ of small, asymptotically trivial transformations  
combines
the group $\cald_0$ of infinitesimally generated, asymptotically
trivial diffeomorphisms  with asymptotically trivial
Chern-Simons  transformations.
In terms of the restriction $\rho_0$ of $\rho_\infty$ to
 $\cald_0 \ltimes \calp_0 $
we have
\bea
\label{smalltrans}
\calt_0=(\cald_0 \ltimes \calp)/\mbox{ker}\;\rho_0,
\eea
where we have again used that fact that, due to the contractibility
of $\PPoi$, all  elements of $\mathcal P$ are connected to the
identity.

We are now ready to determine the asymptotic symmetry group $\calb$ and the
symmetry group $\cals$ of $\PPoi$-Chern-Simons theory as defined in
\eqref{bundle}. Using 
$\mbox{ker}\;\rho=\mbox{ker}\;\rho_\infty\cap \calt$ 
and $\mbox{ker}\;\rho_0=\mbox{ker}\;\rho_\infty \cap \calt_0$ 
and standard results for quotients involving normal 
subgroups (see  for example
the book \cite{lang} by Lang), it follows that these groups are given by
\begin{align}
&\calb=\calt_\infty/\calt = \frac{(\cald_\infty\ltimes\calp_\infty)/
(\cald\ltimes\calp)}{\text{ker}\;\rho_\infty /\text{ker}\;\rho}\label{bg}\\ 
&\cals=\calt_\infty/\calt_0  =\frac{(\cald_\infty\ltimes\calp_\infty)/
(\cald_0\ltimes
 \calp)}{\text{ker}\;\rho_\infty/\text{ker}\;\rho_0}\;.\label{sg}
\end{align}
In order to evaluate these quotients we have to clarify the
relationship between the kernels of the maps $\rho_\infty,\rho$ and
$\rho_0$. We claim
\begin{lemma}
For the maps  $\rho_\infty,\rho$ and
$\rho_0$ defined in  \eqref{inftrans},\eqref{trans} and
\eqref{smalltrans} we have
\bea
\label{ker0}
\text{ker}\;\rho=
\text{ker}\;\rho_0\\
\label{kerinfty}
 \text{ker}\;\rho_\infty /\text{ker}\;\rho \cong \tilde U(1).
\eea
\end{lemma}

{\bf Proof} 
\begin{enumerate}
\item Clearly, $\text{ker}\;\rho_0 \subseteq \text{ker}\;\rho$, but to prove
 $\text{ker}\;\rho_0 = \text{ker}\;\rho$
we need to show that it is impossible for $(\phi,\gamma)$
to be in $\text{ker}\;\rho$ if $\phi$ is a large diffeomorphism.
However, this follows from the explicit determination of the
action of the mapping class group on the holonomies we give 
in Sect.~5.3. Suppose $\phi$ is a large
diffeomorphism and $A \in \cala_\infty$. Computing the holonomies of
$\phi_*A$ around the generators $m_1,...m_n,a_1,b_1,...a_g,b_g$ 
shown in Fig.~3 is equivalent to computing the
holonomies of $A$ around the transformed generators
 $\phi^{-1} \circ m_1$ etc.
However, we show explicitly in Sect.~5.3 that  non-trivial
elements of the mapping
act non-trivially  on the holonomies. On the other hand, gauge transformations
which are $1$ at the boundary act trivially. Thus  no such pair
$(\phi,\gamma)$ can be in $\text{ker}\;\rho$.

\item 
To establish \eqref{kerinfty}
we show  that the evaluation map
  $\text{ker}\;\rho_\infty \rightarrow \tilde U(1)$,
which assigns to each element $(\phi,\gamma) \in
\text{ker}\;\rho_\infty$ the $\tilde U(1)$ rotation in $\gamma(\infty)$
is a surjective homomorphism with kernel $\text{ker}\;\rho$.
It is clear that the map is a homomorphism; to see that it  is surjective
recall the special form   of any connection in $\cala_\infty$
in a neighbourhood of the boundary.  
Equations \eqref{dreib} and \eqref{gaugefield} imply that
any gauge transformation $\gamma_\alpha$ that reduces to a rotation of
angle $-\alpha$ near the boundary
 can
be compensated by a diffeomorphism $\phi_{\{\alpha\}}$ 
which asymptotically is a global rotation
by the mod $2\pi$
reduction  $\{\alpha\}\in[0,2\pi)$ of $\alpha$.
Hence  the pair $(\phi_{\{\alpha\}},\gamma_\alpha)$  
is in $\text{ker}\;\rho_\infty $
for any $\alpha \in \RR$.$\Box$
\end{enumerate}

It seems very plausible that in fact $\text{ker}\;\rho=
\text{ker}\;\rho_0\cong {1}$ and $\text{ker}\;\rho_\infty \cong\tilde U(1)$.
We have not been able to show this rigorously, but for the following
discussion the relations \eqref{ker0} and \eqref{kerinfty} are sufficient.

Using  evaluation maps at infinity analogous to that used in
 the proof of Lemma~5.1 it is easy to check that
$\cald_\infty/\cald\cong U(1)$ and
$\calp_\infty/\calp \cong \tilde
U(1)\times\RR$. Thus we find that the asymptotic symmetry group is
given by
\bea
\calb= \frac{(\cald_\infty\ltimes\calp_\infty)/
(\cald\ltimes\calp)}{\text{ker}\;\rho_\infty /\text{ker}\;\rho}
=\frac{U(1)\times \tilde U(1) \times \RR}{\tilde U(1)}
= U(1)\times \RR.
\eea

Finally, we want to determine the symmetry group $\cals$, which
requires evaluation of the quotient
\bea
\cals^D=\cald_\infty/\cald_0.
\eea 
We know that
$\cald/\cald_0=\text{Map}(S^\infty_{g,n})$, and first give a
 heuristic description\footnote{We thank
   D. Giulini, for pointing out this
 part of the argument to CM} of $\cals^D$, which will then be justified by
 a rigorous proof. Consider the effect of a diffeomorphism which
is a rotation at the boundary on the generators of
the  fundamental group. If the angle of the rotation exceeds $2\pi$,
the diffeomorphism wraps the curves representing the generators
around the boundary at infinity as depicted in Fig.~4.

\vbox{
\vskip .3in
\input epsf
\epsfxsize=12truecm
\centerline{
\epsfbox{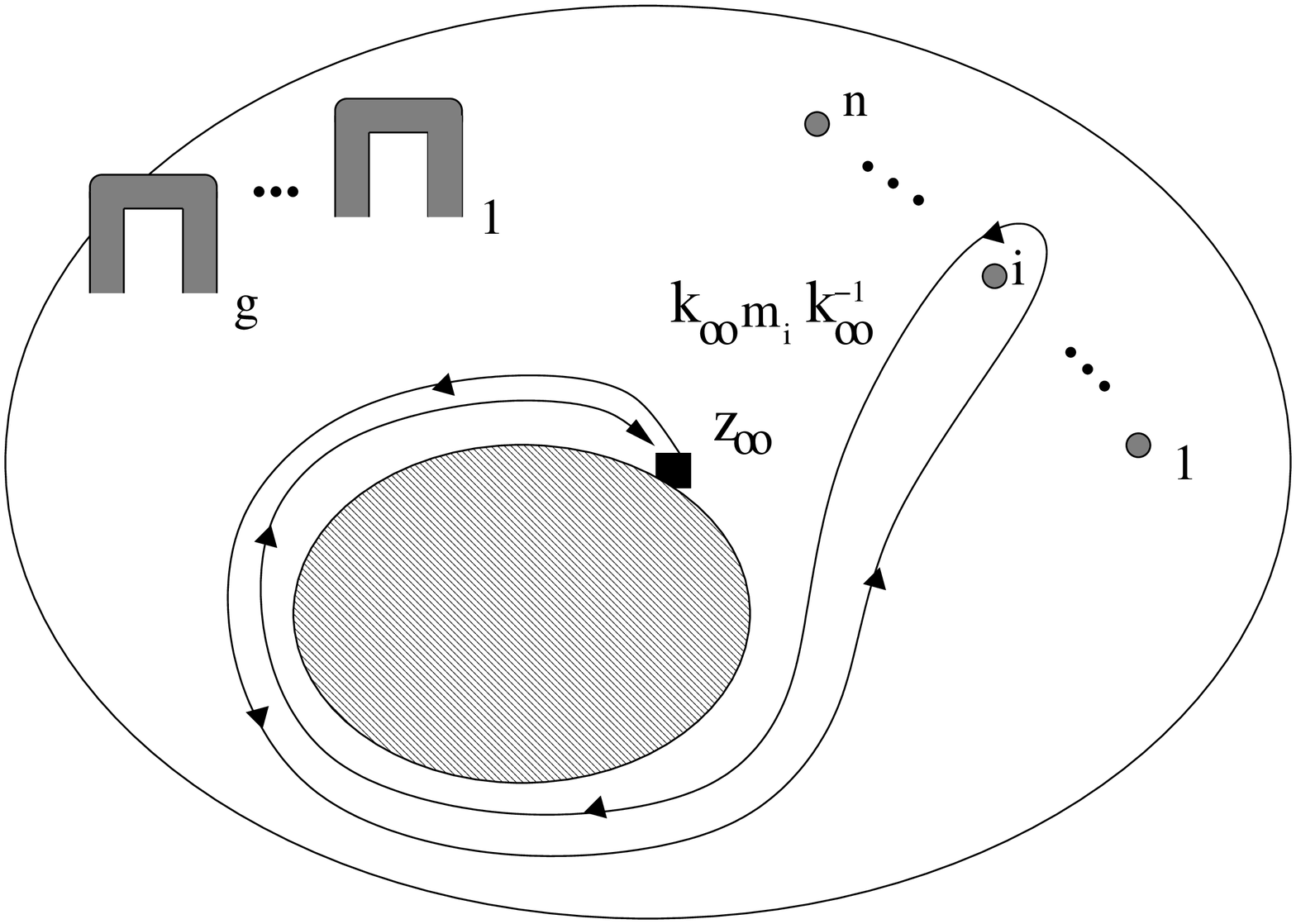}
}
\bigskip
{
\centerline{\bf Fig.~4 }
}
\centerline{\footnotesize The action of a rotation at the boundary}
\bigskip}

It then acts nontrivially on the  fundamental
group by conjugating every element  with the element
\bea
\label{infhol}
k_\infty=b_ga_g^{-1}b_g^{-1}a_g\cdots b_1 a_1^{-1}b_1^{-1}a_1m_n\cdots m_1.
\eea
 We can thus identify it 
with a combination of a large, asymptotically trivial diffeomorphism
$C_\infty$ acting on the fundamental group by global conjugation
with $k_\infty$ and a small diffeomorphism that reduces to a
asymptotic rotation by  $\alpha-2\pi$. 
Note that the element  $[C_\infty]\in \text{Map}(S^\infty_{g,n})$ 
is central, which  follows from the
fact that the equivalence class of the  path $k_\infty$ around
the boundary is invariant under diffeomorphisms that keep the
boundary fixed. Denoting by $\Omega$ the rotation
 by $2\pi$ in the
cover $\tilde U(1)$ and writing $\ZZ$ for the central subgroup of
$\text{Map}(S^\infty_{g,n})\times\tilde U(1)$ generated by
$([C_\infty],\Omega)$, we claim 
\begin{lemma}
The quotient $\cals^D=\cald_\infty/\cald_0$ is given by
\bea
\label{dgroup}
\cals^D=(\mbox{\rm Map}(S^\infty_{g,n})\times\tilde
  U(1))
/\ZZ.
\eea
\end{lemma}
{\bf Proof} Our proof closely
follows the argument given in \cite{giu1}, \cite{giu2}. We consider
the non-trivial part of the exact sequence associated to the bundle structure
\bea
\label{bbundle}
\begin{array}{ccc}
\text{Map}(S^\infty_{g,n})\cong \cald/\cald_0 & \xrightarrow{i} &
\cald_\infty/\cald_0 \cong \cals^D\\
& & \downarrow p\\
& &\cald_\infty/\cald\cong U(1),
\end{array}
\eea
which takes the form
\bea
\label{sequence2}
1\rightarrow \pi_1(\cald_\infty/\cald_0)\xrightarrow{p_*}\pi_1(U(1))\cong
\mathbb{Z}
\xrightarrow{\partial_*}\text{Map}(S^\infty_{g,n})
\xrightarrow{i_*}\cals^D/\cals^D_0\rightarrow 1,
\eea
where the function $\partial_*$ is obtained as follows. We parametrise
$\cald_\infty/\cald \cong U(1)$ by $e^{i \alpha }$,
$\alpha \in [0,2\pi)$ and choose the loop
$\Gamma(\alpha )=e^{i\alpha}$, $\alpha \in[0,2\pi]$, whose homotopy
class generates $\pi_1(U(1))\cong\mathbb{Z}$. By assigning to
every element $e^{i\alpha}\in U(1)$ the
equivalence class of
diffeomorphisms which are a global
rotation of angle $\alpha$ at spatial infinity,
we can define a curve
$\bar{\Gamma}$ in $\cals^D=\cald_\infty/\cald_0$
that starts at the identity. The projection
$p_*$ in \eqref{sequence2} maps an equivalence class of diffeomorphisms
in $\cals^D$
to the value they take at spatial infinity. So we see that
$p_*\circ\bar{\Gamma}=\Gamma$, and $\bar{\Gamma}$ is the lift of
$\Gamma$ to $\cals^D$. Visualising the action of the mapping classes
$\bar{\Gamma}(\alpha)$ by their action on a set of curves representing the
generators of the fundamental group, we see that the curves are
wrapped once around the boundary at spatial infinity as $\alpha$ approaches
$2\pi$ (Fig.~4). Therefore, we have
$\partial_*([\Gamma])=\bar{\Gamma}(2\pi)=[C_\infty]$. It follows from the
exactness of \eqref{sequence2} that
$\pi_1(\cals^D)=\{1\}$ and $\cals^D/\cals^D_0\cong\mathbb{Z}$.
This means that the holonomy group of the bundle \eqref{bbundle}
is the central subgroup of 
$\text{Map}(S^\infty_{g,n}) $ generated by $[C_\infty]$
and that $\cals^D$ is given by \eqref{dgroup}.$\Box$

Using Lemmas 5.1 and 5.2, we can evaluate the
quotients occuring in \eqref{sg} and obtain
symmetry group
\bea
\label{symgroup}
\cals
= \frac{\left((\text{Map}(S^\infty_{g,n})
\times\tilde U(1))/\ZZ \right)\times \tilde U(1)\times\RR}{\tilde U(1)}
=\frac{\text{Map}(S^\infty_{g,n})
\times\tilde U(1) \times\RR}{\ZZ}.
\eea

The above discussion can  be generalised  from the
centre of mass frame to a general frame. As explained  at the
beginning of Sect.~4, this is easily accommodated in the Chern-Simons
 formulation. We then allow general Poincar\'e transformations
as asymptotic symmetries. Writing again  $\ZZ$ for the central subgroup of
$\text{Map}(S^\infty_{g,n})\times \PPoi$ generated by
$([C_\infty],\exp(-2\pi J_0))$,
the  symmetry  group is the quotient
\bea
\label{symgroupp}
\frac{\text{Map}(S^\infty_{g,n})\times \PPoi}{\ZZ}.
\eea

\subsection{Hamiltonian, total angular momentum and conserved quantities}

 After determining the symmetry group of our model,
we now investigate how the continuous symmetries at spatial
infinity are related to conserved quantities.
 We show that the total momentum and total angular momentum
 three-vector of the universe generate asymptotic Poincar\'e
transformations.  In the centre of mass frame of the universe,
 only spatial rotations and time translations are
admitted. The conserved quantities associated to these
transformations are the total angular momentum and the
total energy of the universe.

The total momentum and angular momentum three-vectors of the universe
as seen by the observer are given by the holonomy $h_\infty$
of the curve $k_\infty$ \eqref{infhol} around the boundary pictured in
Fig.~3
\bea
\label{infprod}
h_\infty&=&(u_{tot},\ba_{tot})=(\exp(-p_{tot}^aJ_a),
-\Ad(\exp(-p_{tot}^a J_a))\bj_{tot})\\
&=&({B_g}{A_g}^{-1}{B_g}^{-1}{A_g})
\cdots({B_1}{A_1}^{-1}{B_1}^{-1}{A_1}){M_n}\cdots {M_1},\nonumber
\eea
which gives
\bea
\label{utot}
u_{tot}&=&\exp(-p_{tot}^aJ_a)=u_{K_g}\cdots u_{K_1} u_{M_n}\cdots u_{M_1}\\
\bj_{tot}&=&\sum_{i=1}^n \Ad({u_{M_1}^{-1}\cdots
  u_{M_{i-1}}^{-1}})\bj^{M_i}
\label{jtot}\\
&+&\sum_{i=1}^g \bigg(\;\Ad({ u_{M_1}^{-1}\cdots
  u_{M_n}^{-1}u_{K_1}^{-1}\cdots u_{K_i}^{-1}  })\big(
(\Ad({u_{B_i}})-\Ad({u_{B_i}u_{A_i}^{-1}}))\bj^{B_i}\nonumber\\
&\quad&\qquad-\Ad({ u_{M_1}^{-1}\cdots
  u_{M_n}^{-1}u_{K_1}^{-1}\cdots u_{K_i}^{-1}
  })\big((\Ad({u_{B_i}})-\Ad({u_{K_i}}))
\bj^{A_i}\bigg)\nonumber
\eea
with $u_{K_j}:=u_{B_j}u_{A_j}^{-1}u_{B_j}^{-1}u_{A_j}$ for
$j=1,\ldots,g$. In general, the element $u_{tot}\in\tilde
L_3^\uparrow$ does not have to be elliptic, but can also be parabolic
or hyperbolic, which gives rise to interesting effects such as
Gott-pairs \cite{Gott}. In this article, we
will restrict our interpretation to the elliptic case and assume
$p^0_{tot}>0$, $\bp_{tot}^2=\mu^2$ with $\mu\in(0,2\pi)$.
We see that the total momentum and angular momentum
three-vectors $\bp_{tot}$, $\bj_{tot}$ are {\em not}
obtained by simply adding the corresponding quantities for the
individual handles and particles. However, using the decoupled
coordinates \eqref{pjseptrafo} defined in Sect.~\ref{phasespace},
the angular momentum of the universe can be rewritten as
\bea
\label{jtotdecoup}
\bj_{tot}&=&\sum_{i=1}^n \bj^{M'_i}+\sum_{i=1}^g \bj^{H'_i}
\eea
where $\bj^{H'_i}$ is given by equation \eqref{jhandle} and can be interpreted as the angular momentum associated to the $i^{th}$ handle.

In order to calculate the Poisson brackets involving these quantities,
we must determine how right and left multiplication of the
generators of the fundamental group affect the
total holonomy \eqref{infprod}. If we define
\begin{align}
&\Psi:\mathcal{C}_1\times\ldots\mathcal{C}_n\times
\left(\PPoi\right)^{2g}\rightarrow \PPoi\times\mathcal{C}_1
\times\ldots\mathcal{C}_n\times \left(\PPoi\right)^{2g}\\
&\Psi({M_1},\ldots,{B_g})=(h_\infty,{M_1},\ldots,{B_g}),\nonumber
\end{align}
the derivative of the map $\Psi$ induces
a map of the vector fields associated to the generators.
Expressing the image of the right and left invariant
vector fields on $\mathcal{C}_1\times\ldots\mathcal{C}_n\times \left(\PPoi\right)^{2g}$
in terms of the right and left invariant vector fields
on $\PPoi\times\mathcal{C}_1\times\ldots\mathcal{C}_n\times \left(\PPoi\right)^{2g}$ and
inserting them in the Poisson bivector \eqref{fundbivect},
we obtain the image of the Poisson bivector
\bea
\label{totfundbivect}
\Psi_*(B_{FR})&=B_{FR}+\tfrac{1}{2}r^{\alpha\beta}(L^{tot}_\alpha\wedge
L^{tot}_\beta+R^{tot}_\alpha\wedge R^{tot}_\beta)+r^{\alpha\beta}
R^{tot}_\alpha\wedge L^{tot}_\beta\\
&+r^{\alpha\beta}R^{tot}_{\alpha}\wedge\left(\left(\sum_{i=1}^n R^\mi_\beta+
L^\mi_\beta\right)+\left(\sum_{i=1}^g R^\ai_\beta+L^\ai_\beta+
R^\bi_\beta+L^\bi_\beta\right)\right)\nonumber\\
&+r^{\alpha\beta}\left(\left(\sum_{i=1}^n R^\mi_\alpha+
L^\mi_\alpha\right)+\left(\sum_{i=1}^g R^\ai_\alpha+L^\ai_\alpha+
R^\bi_\alpha+L^\bi_\alpha\right)\right)\wedge L^{tot}_\beta\nonumber,
\eea
where $L^{tot}$ and $R^{tot}$ are the right and
left invariant vector fields associated to the
element $h_\infty$. With the $r$-matrix $r=P_a\otimes J^a$
and expressions \eqref{vecfields} for the right and
left invariant vector fields, we derive the Poisson brackets
\begin{align}
\label{totalbr}
&\{j_{tot}^a, j_{tot}^b\}=
-\epsilon^{abc}j^{tot}_c\qquad\{p_{tot}^a, j_{tot}^b\}=
-\epsilon^{abc}p^{tot}_c\qquad\{p_{tot}^a, p_{tot}^b\}=0 \\
\nonumber\\
\label{jbr}
&\{j_{tot}^a,p_X^b\}=-\epsilon^{abc}p_c^X\\
&\{j_{tot}^a,j_X^b\}=-\epsilon^{abc}j_c^X\nonumber\\
\nonumber\\
\label{pbr}
&\{p_{tot}^a,p_X^b\}=0\\
&\{p_{tot}^a,j_X^b\}=
-\left((1-\Ad(u_X^{-1}))\cdot\frac{1}
{1-\Ad(u_{tot}^{-1})}\right)^{b}_{\;\;\;c}\epsilon^{acd}
p^{tot}_d=
(1-\Ad(u_X^{-1}))^b_{\;\;c}(T^{-1}(\bp_{tot}))^{ca},\nonumber
\end{align}
where $X\in\{M_1\ldots,B_g\}$ and $T(\bp_{tot})$
is given
by equation \eqref{Ttrans}, its inverse by \eqref{tdef}.
We see that the Poisson brackets of the components of the
total momentum and the total angular momentum
three-vector are those of a free massive particle,
as required. Together with the brackets \eqref{jbr}
they confirm the interpretation of $\bj_{tot}$ as
the universe's total angular momentum three-vector.

Furthermore, the brackets \eqref{jbr} show that the
components of $\bj_{tot}$ are the infinitesimal
generators of asymptotic Poincar\'e transformations.
Boosts and rotations of the universe with respect
to the reference frame of the observer
act on the group elements associated to
the particles and handles via global conjugation.
For an infinitesimal boost or rotation we get
\begin{align*}
&(u_X,\ba_X)\rightarrow (\exp(\delta_aJ^a),0)(u_X,\ba_X)
(\exp(-\delta_a J^a),0)\\
\\
\Rightarrow\qquad &\delta p_X^b=\delta^a(J_a)^b_{\;\;c}p_X^c=
-\delta_a\epsilon^{abc} p^X_c=\delta_a\{j^a_{tot},p^b_X\}\\
&\delta j_X^b=\delta^a(J_a)^b_{\;\;c}j_X^c=-\delta_a\epsilon^{abc}
j^X_c=
\delta_a\{j^a_{tot},j^b_X\}.
\end{align*}
The interpretation of the action \eqref{pbr} of the total
momentum three-vector $\bp_{tot}$ is less obvious. It should be
expected that its components generate infinitesimal translations
of the matter in the universe with respect to the observer.
However, it turns out that these are not
translations of the position coordinate $\bx$ but related to the
alternative position coordinate $\bq$ \eqref{postrans}
introduced in Sect.~\ref{oneparticle}. A translation by
$\by$  acts on the group
elements of particles and handles via
\begin{align*}
&(u_X,\ba_X)\rightarrow(1,\by)(u_X,\ba_X)(1,-\by)\\
\\
\Rightarrow\qquad &\delta p^b_X=0\\
 &\delta j_X^b=(1-\Ad(u_X^{-1}))^{bc}y_c,
\end{align*}
and by comparison with \eqref{pbr} we get
\begin{align*}
&\delta p^b_X=\delta_a\{p_{tot}^a,p^b_X\}\\
 &\delta j_X^b=\delta_a\{p_{tot}^a,j^b_X\} &
&\text{with}\;y^a=\left(T^{-1}(\bp_{tot})\right)^{ab}\delta_b.
\end{align*}
Recalling the definition \eqref{postrans} of the alternative position
 coordinate $\bq$, we see that this corresponds to the infinitesimal
 change of $\bx$ induced by an infinitesimal change in $\bq$
\bea
\label{xtrans}
q^a&\rightarrow& q^a+\delta^a\\
x^a&\rightarrow& x^a+\left(T^{-1}(\bp)\right)^a_{\;\;b}\delta^b,
\eea
only that now the map $T(\bp_{tot})$ occurring in
the formula corresponds to the total
momentum three-vector $\bp_{tot}$ of the universe instead of the
momentum three-vector of the particle. Note
that for an infinitesimal translation in the direction of the total
three-momentum $\bp_{tot}$ we have $\delta^a=y^a$ and get the familiar
 transformation properties.

The time-component of the three-vector $\bp^{tot}$ has
the interpretation of the total energy of the universe as
 measured by the  observer. It is the
Hamiltonian of the system\footnote{In a system with
  constraints, the Hamiltonian is only determined
  up to a linear combination of the constraints. However, as
  the constraints \eqref{constraint2} Poisson commute with all
  functions on the phase space, their
  contribution is trivial and can be omitted.} and generates
its time development
\bea
H=p_{tot}^0.
\eea As it commutes with all three-momenta, the total three momentum
  of the universe as well as all individual three-momenta of particles and
  handles are conserved quantities. This reflects the fact that the
  interaction in (2+1)-dimensional gravity is topological, not
  dynamical in nature.  Other conserved quantities are the
  zero-component of
the total angular momentum three-vector and the total mass $\mu$
and spin $s$ of the universe, given by
\bea
\bp_{tot}^2=\mu^2\qquad\qquad\bp_{tot}\cd \bj_{tot}=\mu s.
\eea

The reference frame in which the centre of mass of the
  universe is at rest at the origin is characterised by the conditions
\bea
\label{restframe}
\bp_{tot}=(H,0,0)\qquad\text{and}\qquad \bj_{tot}=(j_{tot}^0,0,0),
\eea
where $H=\mu$ is the total energy or mass of the universe
and $j_{tot}^0=s$ its total angular momentum or spin.
The Hamiltonian is the conserved quantity associated
to  translations in the time direction.
The spin of the universe, which can easily be seen to
Poisson commute with the Hamiltonian, is the conserved
quantity associated to asymptotic spatial rotations.
The Hamilton equations for the total momentum and angular
momentum three-vector in the momentum rest frame of the
universe are obtained from equation \eqref{totalbr}
\bea
\label{hamequ}
\dot{p}_{tot}^a=\{H,p_{tot}^a\}=0\qquad\dot{j}_{tot}^a=\{H,j_{tot}^a\}=0.
\eea
As expected, in the reference frame where the centre of mass of
the universe is at
rest at the origin, the universe's total momentum and total angular
momentum three-vector have no time development and stay of the form
\eqref{restframe}.

\subsection{The action of the mapping class group}

We know from Sect.~4.2 that
the elements of the mapping class group
$\text{Map}(S^\infty_{g,n})$ act on
the phase space \eqref{npartphspace}
via automorphisms of the fundamental group.
However, we have yet to
determine how this action affects the Poisson structure.

For explicit calculations we need to choose a set of generators
of the mapping class group. This is best done in two steps.
First we consider the pure mapping class group
$\text{PMap}(S^\infty_{g,n})$ which, by definition,
is the  subgroup of $\text{Map}(S^\infty_{g,n})$ which leaves each puncture
fixed. As explained in \cite{birmgreen}, it
 is generated by Dehn twists along a set of curves on the
 surface. For our purposes it is most convenient to
 work with the set of generators given
 by Schomerus and Alekseev \cite{AS}.
This set is obtained from the homotopy classes of the curves $a_i$, $\delta_i$, $\alpha_i$, $\epsilon_i$
 and $\kappa_{\nu\mu}$ pictured in Fig.~5.

\vbox{
\vskip .3in
\input epsf
\epsfxsize=12truecm
\centerline{
\epsfbox{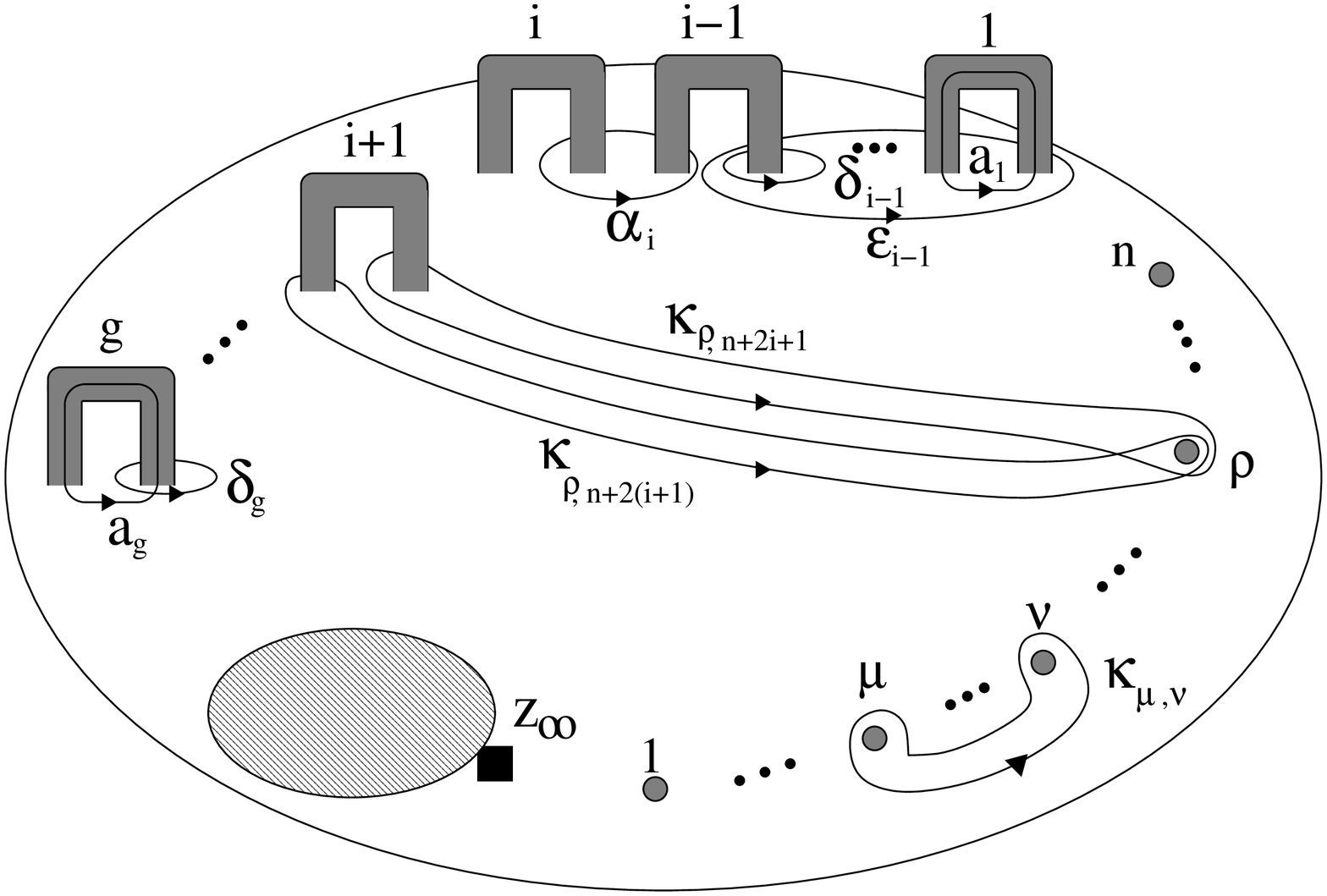}
}
\bigskip
{
\centerline{\bf Fig.~5 }
}
\centerline{\footnotesize The curves associated to the generators of the (pure)
  mapping class group}
\bigskip}

In terms of the generators of the fundamental group they are given by
\begin{align}
\label{twistcurves}
&a_i &  &i=1,\ldots,g\\
&\delta_i=a_i^{-1}b_i^{-1}a_i & &i=1,\ldots,g\nonumber\\
&\alpha_i=a_i^{-1}b_i^{-1}a_ib_{i-1} & &i=2,\ldots,g\nonumber\\
&\epsilon_i=a_i^{-1}b_i^{-1}a_i
\cdot(b_{i-1}a^{-1}_{i-1}b^{-1}_{i-1}a_{i-1})
\cdot\ldots\cdot(b_1a_1^{-1}b_1^{-1}a_1)
& &i=2,\ldots,n\nonumber\\
&\kappa_{\nu,\mu}=m_\mu m_\nu & &1\leq \nu<\mu\leq n\nonumber\\
&\kappa_{\nu, n+2i-1}=a_i^{-1}b_i^{-1}a_i m_\nu & &\nu=1,
\ldots,n,\,i=1,\ldots,g\nonumber\\
&\kappa_{\nu, n+2i}=b_i m_\nu & &\nu=1,\ldots,n,\,i=1,\ldots,g\;.\nonumber
\end{align}
Define a Dehn twist around each of these curves by embedding a small
annulus around the curve and twisting its ends by an angle of $2\pi$
as shown in Fig.~6.

\vbox{
\vskip .3in
\input epsf
\epsfxsize=12truecm
\centerline{
\epsfbox{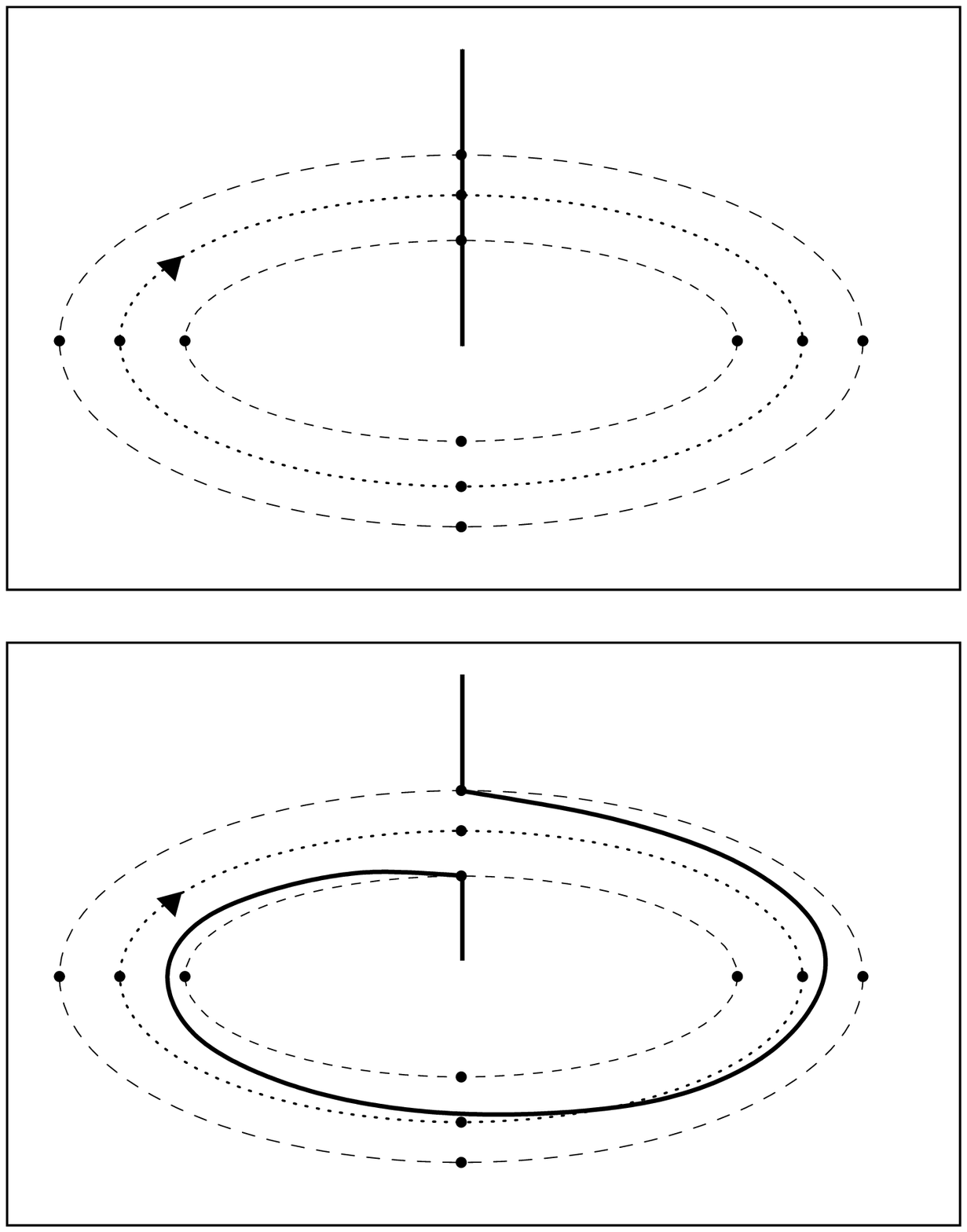}
}
\bigskip
{
\centerline{\bf Fig.~6 }
}
\centerline{\footnotesize The effect of a 
 Dehn twist around an oriented loop (dotted
  line)}
\centerline{\footnotesize on  a curve intersecting the loop 
transversally (full line)}
\bigskip}

 A Dehn twist around a given curve
then defines an outer automorphism of the fundamental group, affecting
only those elements whose representing curves intersect with
it. Drawing the images of these curves as indicated in Fig.~6,
we obtain  the action of the pure mapping class group on the
fundamental group of the surface. The associated action on the
holonomies is given by the following table, where for each
transformation we list only the holonomies that do not transform trivially
\begin{align}
\label{ad}
a_i: &B_i \rightarrow B_iA_i\\
\nonumber\\
\label{dd}
\delta_i: &A_i \rightarrow B_iA_i\\
\nonumber\\
\label{alpha}
\alpha_i: &A_i\rightarrow B_i^{-1}A_iB_{i-1}=A_i\alpha_i\\
&B_{i-1}\rightarrow
B_{i-1}^{-1}A_i^{-1}B_iA_iB_{i-1}A_i^{-1}B_i^{-1}A_iB_{i-1}=
\alpha_i^{-1}B_{i-1}\alpha_i\nonumber\\
&A_{i-1}\rightarrow B_{i-1}^{-1}A_i^{-1}B_iA_iA_{i-1}=
\alpha_i^{-1}A_{i-1}\nonumber\\
\nonumber\\
\label{eps}
\epsilon_i: &A_i\rightarrow B_i^{-1}A_iK_{i-1}\ldots K_1=A_i\epsilon_i
\qquad\qquad K_j:=B_j A_j^{-1}B_j^{-1}A_j\nonumber\\
&A_k\rightarrow K_1^{-1}\ldots K_{i-1}^{-1}(A_i^{-1}B_iA_i) A_k
(A_i^{-1}B_i^{-1}A_i)K_{i-1}\ldots K_1=\epsilon_i^{-1}
A_k\epsilon_i & &\forall 1\leq k
<i\\
&B_k\rightarrow K_1^{-1}\ldots K_{i-1}^{-1}(A_i^{-1}B_iA_i) B_k
(A_i^{-1}B_i^{-1}A_i)K_{i-1}\ldots K_1=\epsilon_i^{-1}B_k\epsilon_i &
&\forall 1\leq k <i\nonumber\\
\nonumber\\
\label{etapp}
\kappa_{\nu\mu}: &M_\nu\rightarrow M_\nu^{-1}M_\mu^{-1}M_\nu M_\mu
M_\nu=\kappa_{\nu\mu}^{-1}M_\mu\kappa_{\nu\mu}\\
&M_\mu\rightarrow M_\nu^{-1}M_\mu
M_\nu=\kappa_{\nu\mu}^{-1}M_\mu\kappa_{\nu\mu}\nonumber\\
&M_\kappa\rightarrow  M_\nu^{-1}M_\mu^{-1}M_\nu M_\mu M_\kappa M_\mu^{-1}
M_\nu^{-1} M_\mu M_\nu & &\forall \nu<\kappa<\mu\nonumber\\
\nonumber\\
\label{etapdelta}
\kappa_{\nu, n+2i-1}: &M_\nu\rightarrow
M_\nu^{-1}A_i^{-1}B_iA_iM_\nu A_i^{-1}B_i^{-1}A_iM_\nu=
\kappa_{\nu,n+2i-1}^{-1}M_\nu\kappa_{\nu, n+2i-1}\\
&A_i\rightarrow B_i^{-1}A_iM_\nu=A_i\kappa_{\nu,n+2i-1}\nonumber\\
& X_j\rightarrow M_\nu^{-1}A_i^{-1}B_iA_iM_\nu A_i^{-1}B_i^{-1}A_i
X_jA_i^{-1}B_iA_iM_\nu^{-1}A_i^{-1}B_i^{-1}A_i
M_\nu\nonumber\\
&\qquad=\kappa_{\nu,n+2i-1}^{-1}M_\nu\kappa_{\nu,n+2i-1}M_\nu^{-1}X_j M_\nu
\kappa_{\nu,n+2i-1}^{-1}M_\nu^{-1}\kappa_{\nu,n+2i-1}, \nonumber\\
&\qquad\qquad X_j\in\{M_{\nu+1},\ldots,M_n,A_1,\ldots,B_{i-1}\}\nonumber
\end{align}
\begin{align}
\label{etapb}
\kappa_{\nu,n+2i}: &M_\nu\rightarrow M_\nu^{-1}B_i^{-1}M_\nu
B_iM_\nu=\kappa_{\nu,n+2i}^{-1}M_\nu\kappa_{\nu,n+2i}\\
 &B_i\rightarrow
 M_\nu^{-1}B_iM_\nu=\kappa_{\nu,n+2i}^{-1}B_i\kappa_{\nu,n+2i}\nonumber\\
&A_i\rightarrow M_\nu^{-1}B_i^{-1}A_i B_i^{-1}M_\nu^{-1}B_i
M_\nu\nonumber\\
&\qquad=\kappa_{\nu,n+2i}^{-1}A_iB_i^{-1}M_\nu^{-1}\kappa_{\nu,n+2i}\nonumber\\
&X_j\rightarrow
M_\nu^{-1}B_i^{-1}M_\nu B_iX_jB_i^{-1}M_\nu^{-1}B_i
M_\nu\nonumber\\&\qquad=\kappa_{\nu,n+2i}^{-1}M_\nu B_iX_jB_i^{-1}
M_\nu^{-1}\kappa_{\nu,n+2i},
& &X_j\in\{M_{\nu+1},\ldots,M_n, A_1,\ldots B_{i-1}\} \nonumber.
\end{align}
The full mapping class group of the surface $S_{g,n}^\infty$ is
related to  pure mapping class group by
the short exact sequence
\bea
1\rightarrow\text{PMap}(S^\infty_{g,n})\xrightarrow{i}
\text{Map}(S^\infty_{g,n}) \xrightarrow{\pi}
S_n \rightarrow 1,
\eea
where $i$ is the canonical  embedding and $\pi$ is the projection onto the
permutation group defined in \eqref{permute}.
It follows  that we  obtain a set of generators for the
full mapping class group by supplementing the set of generators of the
pure mapping class group with a set of  generators $\sigma_i$
  which get mapped to the elementary
transpositions
via $\pi$. As explained in \cite{birmgreen},\cite{birmorange}
the $\sigma_i$ generate  the braid group on the
surface $S_{g,n}^\infty$.
The action of these generators on the holonomies around the punctures
is shown in Fig.~7.

\vbox{
\vskip .3in
\input epsf
\epsfxsize=12truecm
\centerline{
\epsfbox{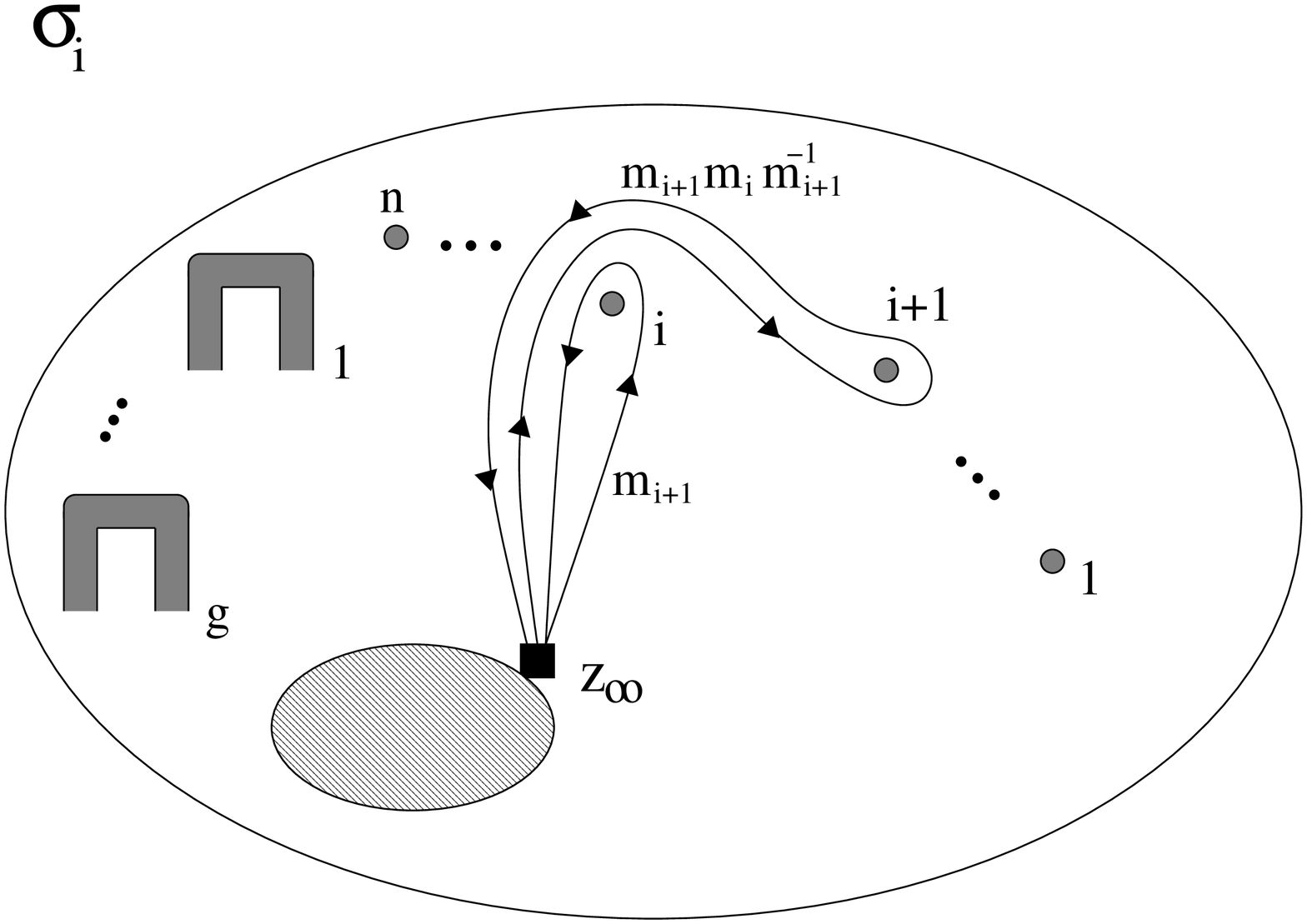}
}
\bigskip
{
\centerline{\bf Fig.~7 }
}
\centerline{\footnotesize Generators of the braid group on the surface $S_{g,n}^\infty$}
\bigskip}

They act on the phase space variables $M_i$ according to
\begin{align}
\label{braid}
\sigma_i:
&M_i\rightarrow M_{i+1}\\
&M_{i+1}\rightarrow M_{i+1}M_iM_{i+1}^{-1}\;.\nonumber
\end{align}
Armed with  the expressions \eqref{ad} to \eqref{braid} for the
action of the (full) mapping class group on the generators
of the holonomies  we can prove
\begin{theorem}
The mapping class group $\text{Map}(S^\infty_{g,n})$
acts on the phase space via   Poisson
isomorphisms.\footnote{A similar claim for a
  quantised version of the holonomies was made by Schomerus and
  Alekseev in \cite{AS}. A proof was announced in \cite{AS}, but never published.}
\end{theorem}
\noindent{\bf Proof} The action of the Dehn twists \eqref{ad} to
\eqref{braid} on the holonomies induces a transformation of the
associated left and right invariant vector fields.
This transformation can be calculated explicitly,
allowing one to express the images of the vector fields
in terms of the left and right invariant vector
fields associated to the images of the holonomies.
We demonstrate the general pattern on the case of
the action \eqref{braid} of the generators $\sigma_i$.
Determining how left and right multiplication of the
holonomies affect their images under \eqref{braid},
we obtain an expression for the image of right and
left invariant vector fields in terms of the vector
fields associated to the images of the holonomies:
\begin{align}
\label{image}
R^\mi&\rightarrow\Ad(M_{i+1})R^{\sigma(M_{i+1})}=\Ad(\sigma(\mi))R^{\sigma(M_{i+1})}\\
L^\mi&\rightarrow\Ad(M_{i+1})L^{\sigma(M_{i+1})}=\Ad(\sigma(\mi))L^{\sigma(M_{i+1})}\nonumber\\
R^{M_{i+1}}&\rightarrow R^{\sigma(\mi)}-\Ad(M_{i+1})
(R^{\sigma(M_{i+1})}+L^{\sigma(M_{i+1})})\nonumber\\
&\qquad=R^{\sigma(\mi)}-\Ad(\sigma(\mi))
(R^{\sigma(M_{i+1})}+L^{\sigma(M_{i+1})})\nonumber\\
L^{M_{i+1}}&\rightarrow L^{\sigma(\mi)}+R^{\sigma(M_{i+1})}+L^{\sigma(M_{i+1})}\nonumber\;.
\end{align}
Note that the sum of all the vector fields that are
affected by this transformation is form-invariant
\bea
R^\mi+L^\mi+R^{M_{i+1}}+L^{M_{i+1}}\rightarrow
R^{\sigma(\mi)}+L^{\sigma(\mi)}+R^{\sigma(M_{i+1})}+L^{\sigma(M_{i+1})}\;,
\eea
implying form-invariance of the sum $R^\mi+L^\mi+\cdots+R^{B_g}+L^{B_g}$.
This is not particular to transformation
\eqref{braid} but a general pattern 
for all the transformations \eqref{ad} to \eqref{braid}.

To prove that the Dehn twists act as Poisson isomorphisms, we have to
show that the Poisson bivector \eqref{fundbivect} is
form-invariant. This is most easily seen by writing the Poisson
bivector in the form \eqref{frbivect2}, as a sum of a term tangential
to the gauge orbits, which
depends only on the antisymmetric part of the $r$-matrix, and a
 term transversal to them, involving only on the matrix $t^{\alpha\beta}$
representing the bilinear form. Now recall Fock
and Rosly's description of the moduli space \cite{FR} and interpret
the fundamental group as a graph. Both, the original and the
transformed graph could be used to describe the Poisson structure  on
the moduli space and should therefore be identical up to gauge
transformations, which in this case are realised as global
conjugations at the vertex. However, as  the part of the Poisson bivector
tangential to the gauge orbits is proportional to the 
sum $R^\mi+L^\mi+\cdots+R^{B_g}+L^{B_g}$,
its form-invariance can be seen immediately
from expression \eqref{frbivect2}.

An explicit proof of the form-invariance of the part transversal to
the gauge orbits proceeds as follows.
We replace the vector fields occurring
in the Poisson bivector by their images
\eqref{image} and simplify the resulting expression using the
relations between right and left invariant vector fields and the
Ad-invariance of the Matrix $t^{\alpha\beta}$ representing the bilinear form.
In the case of the transformation \eqref{braid}, this yields
\begin{align*}
&t^{\alpha\beta}\bigg( R^\mi_\alpha \wedge L^\mi_\beta +
R^{M_{i+1}}_\alpha \wedge L^{M_{i+1}}_\beta +
(R^\mi_\alpha+L^\mi_\alpha )
\wedge ( R^{M_{i+1}}_\beta+L^{M_{i+1}}_\beta ) \bigg)
\rightarrow\\
&t^{\alpha\beta}
\bigg(\left(\Ad(\sigma(\mi))R^{\sigma(M_{i+1})}\right)_\alpha
\wedge\left(\Ad(\sigma(\mi))L^{\sigma(M_{i+1})}\right)_\beta \\
&\qquad\qquad\qquad\qquad\qquad\qquad
\qquad\qquad+
R^{\sigma(\mi)}_\alpha\wedge\left(
  L^{\sigma(\mi)}+R^{\sigma(M_{i+1})}+L^{\sigma(M_{i+1})}\right)_\beta
\\
&\qquad\qquad\qquad\qquad\qquad\qquad
\qquad\qquad+
\left(\Ad(\sigma(\mi))(R^{\sigma(M_{i+1})}+L^{\sigma(M_{i+1})})
\right)_\alpha\wedge R^{\sigma(\mi)}_\beta\bigg)\\
&=t^{\alpha\beta}\left(
R^{\sigma(\mi)}_\alpha\wedge  L^{\sigma(\mi)}_\beta+
R^{M'_{i+1}}_\alpha\wedge L^{M'_{i+1}}_\beta  +
(R^{\sigma(\mi)}_\alpha+L^{\sigma(\mi)}_\alpha)\wedge(R^{\sigma(M_{i+1})}_\beta+
L^{\sigma(M_{i+1})}_\beta ) \right),\nonumber
\end{align*}
demonstrating the form-invariance of the Fock-Rosly bivector.$\qquad\qquad\qquad\qquad\Box$

\section{Conclusion and outlook}
\label{conclusions}
In this paper
 we have  given an explicit description of the phase space of the
 Chern-Simons formulation of
 (2+1)-dimensional gravity
on a  surface
with
with massive, spinning particles  and a
connected boundary component representing spatial infinity. This
description provides a framework in which the physical properties of
the phase space - its Poisson structure, the action of symmetries,
conserved quantities -
can be adressed in a rigorous and systematic way.

Mathematically, our approach is based on  the combinatorial
description of the moduli space of flat connections
developed by Fock and Rosly. However, our main motivation was to give a
description of the phase space of (2+1)
gravity which allows one  to investigate its {\em physical} content.
The various choices we had
to make in constructing our model - the $r$-matrix,  the graph
consisting of generators of the fundamental group,
the treatment of the boundary
conditions - were all determined by that motivation. It turned out
that the  physical requirements could be accommodated
very easily in the Fock-Rosly framework and, conversely,
that the mathematical quantities and concepts of this
framework have a clear  physical interpretation.
Further evidence that our approach is well adapted to the physics of
(2+1) gravity comes from the
fact that it can readily be
extended to include a cosmological constant.

Although we have endeavoured to relate our phase space
to the metric formulation of (2+1) gravity as closely as possible,
the precise relationship
between  the phase space \eqref{npartphspace}
and, for example, the phase space of $n$ particles (but no handles)
discussed from the point of view of the metric formulation in
\cite{Matschull1}
 remains an open and  important question.
As we have indicated in our discussion of the  single particle
in Sect.~3,
it is not difficult to relate the coordinates used in the
two approaches and to show that, for a spinless particle,
our symplectic structure reduces to that found in \cite{MW}.
It would be interesting to check whether our symplectic structure
similarly generalises that found for $n$ spinless particles
in \cite{Matschull1}.

 Understanding the global relationship between
the two phase spaces poses a more difficult challenge.
A key issue here is the invertability of the dreibein, which
is required  in Einstein's
metric formulation  of gravity but not  in the Chern-Simons
formulation.
Matschull argues in  \cite{Matschull2} that this leads to an identification of
states in the Chern-Simons formulation that would not be identified in
the metric formulation, implying different phase spaces for the two
formulations. He illustrates his claim with an example of two
physically distinct spacetimes, which, according to him, should be
identified in the Chern-Simons formulation.
However, an important ingredient of our phase space
\eqref{npartphspace}
is the parametrisation in terms of holonomies together with
a label for the set of curves along which the holonomies are
to be calculated. By insisting that the description of  the same connection
in terms of different sets of generators of the fundamental group
leads to physically distinct states we are able to distinguish
between the spacetimes described in \cite{Matschull2}. In our
formalism,
they are related by an element of the mapping class group (for
particles on a disc without handles this is just the braid group)
and therefore not identified.
By taking sufficient care in the interpretation of our phase
space parameters it thus seems possible to extract spacetime
physics from our formalism which is in agreement with the Einstein's metric
formulation of gravity. One purpose of the present paper
is to provide a firm foundation for such an investigation.

Another important issue to be
adressed in further investigations is the question of
gauge invariant observables. Nelson and Regge studied this question
for closed  surfaces of various genera,  giving a
presentation of the classical algebra of observables and discussing
its quantisation \cite{RN}. Martin investigated the case of a
sphere with an arbitrary number of punctures representing massive
particles with spin
\cite{martin}. By adapting our description of the phase space to the
case of closed
surfaces, we should be able to determine how it is related to these
articles. It would be interesting to see if this allows one
 to treat the case of a
surface without punctures and the punctured sphere in a common framework.

To end, we stress that the description of phase
space based on the work of Fock and Rosly is a
particularly convenient starting point for
quantisation. Fock and Rosly's description of the moduli space of flat
$G$ connections was developped further by Alekseev, Grosse and
Schomerus, who invented a quantisation procedure, the combinatorial
quantisation of Chern-Simons theories, based on this description
\cite{AGSI,AGSII,AS}. In this procedure, which has been worked out
fully for the
case of compact semisimple Lie groups
and applied  to the non-compact group $SL(2,\CC)$
in \cite{BNR}, quantum groups play a key role.
In \cite{bamus}, the relevant quantum group for the Poincar\'e
group and the $r$-matrix \eqref{rmatrix} used in this paper was
identified as the quantum double of the Lorentz group. Moreover, it
was explained how to reconstruct the two-particle Hilbert space
and the scattering cross section of two massive particles
from the representation theory and the
$R$-matrix of the Lorentz double.
This suggests that an extension of this approach
should  enable one  to
quantise the
phase  $n$-massive particles on a genus $g$  surface
discussed in the present paper.

\subsection*{Acknowledgements}
CM thanks D.~Giulini  for
valuable remarks and discussions about the symmetries
of gauge theories on surfaces with boundaries. She acknowledges
financial support by the  Engineering and
Physical Sciences Research Council  and a living stipend from
the Studienstiftung des Deutschen Volkes. BJS thanks H.-J. Matschull
for useful correspondence and acknowledges an Advanced
Research Fellowship of the Engineering
and Physical Sciences Research Council.

\appendix

\section{Classical $r$-matrices and Poisson Lie groups}
\label{poissliefacts}
 This appendix summarises briefly some facts about Poisson-Lie groups
 used in this article. For a more complete and mathematically exact treatment
we refer the reader to the book by Chari and Pressley \cite{CP} and the
articles by Alekseev and Malkin \cite{AMI},\cite{AMII} and
Semenov-Tian-Shansky \cite{setishan}.

\subsection{Lie bialgebras and Poisson-Lie groups}

\subsection*{Poisson-Lie groups and Lie bialgebras}
 A {\em Poisson-Lie group} is a Lie group G that is also a
Poisson manifold such that the multiplication map
$\mu: G\times G\rightarrow G$ is a Poisson map with
respect to the Poisson structure on $G$ and the direct
product Poisson structure on $G\times G$.
Just as a Lie algebra is the infinitesimal object associated to a Lie
group, there is an infinitesimal object,
the {\em Lie bialgebra}, associated to a Poisson-Lie group.
 A { Lie bialgebra}  is a Lie algebra $\frg$ with a skew
symmetric map $\delta:\frg\rightarrow\frg\times\frg$,
the {\em cocommutator}, such that
 $\delta^*:\frg^*\times\frg^*\rightarrow\frg^*$ defines a Lie bracket
 on its dual space $\frg^*$ and $\delta$ is a one-cocycle of
$\frg$ with values in $\frg\otimes\frg$:\bea
\delta([X,Y])=(\ad_X\otimes 1+1\otimes
\ad_X)\delta(Y)-(\ad_Y\otimes 1+1\otimes \ad_Y)
\delta(X)\quad\forall X,Y\in\frg.
\eea
Every Poisson-Lie group $G$ has a unique {\em tangent Lie bialgebra}
$\frg$. Its cocommutator is defined as the dual of the commutator on
$\frg^*$ given by $[\xi_1,\xi_2]_{\frg^*}:
=d_e\{f_1,f_2\}$, where $f_1$, $f_2$ are
 any two smooth functions on $G$ with  $d_ef_1=\xi_1$, $d_e
 f_2=\xi_2\in\frg^*$. (The bracket $[\,,\,]_{\frg^*}$ does not depend
 on the choice of these functions.) Conversely, for every Lie bialgebra
$\frg$ there is a
 unique connected and simply connected Poisson-Lie group with tangent
 Lie bialgebra $\frg$.

\subsection*{Duals and doubles}
As can be inferred from the definitions above, for every Lie bialgebra
$\frg$ its dual space $\frg^*$ is also a Lie
bialgebra. Its commutator is the dual of the cocommutator on $\frg$
and its cocommutator is the dual of the commutator on $\frg$. The
unique
connected and simply connected Poisson-Lie
group $G^*$ with tangent Lie bialgebra $\frg^*$ is
called the {\em dual} of $G$.

A Lie bialgebra $\frg$ and its dual
$\frg^*$ can be combined into a larger Lie bialgebra, its {\em
  classical double}  $\mathcal{D}(\frg)$.
As a Lie algebra, it is the direct sum of the Lie algebra
$\frg$ and its dual $\frg^*$, and its cocommutator is given by
\bea
\label{rcocomm}
\delta_{\mathcal{D}(\frg)}(u)=
(\ad_u\otimes 1+1\otimes\ad_u)(r)\qquad\forall u\in \frg\oplus\frg^*,
\eea
where
$r\in\frg\otimes\frg^*\subset\mathcal{D}(\frg)\otimes\mathcal{D}(\frg)$
is the element associated to the identity map
$\mbox{id}:\frg\rightarrow\frg$ via the canonical pairing of $\frg$
and $\frg^*$. The unique connected and simply connected Poisson-Lie
group $\cald(G)$ with tangent Lie bialgebra $\cald(\frg)$ is called
the {\em double} of the Poisson-Lie group $G$. As a Lie group, the
double is the product of $G$ and its dual $\cald(G)=G\times G^*$. With
respect to a basis $X_\alpha$, $\alpha=1,\ldots, dim\,\frg$, of $\frg$ and the
dual basis $\xi^\alpha$,  $\alpha=1,\ldots,dim\,\frg$, of $\frg^*$, its
Poisson structure is given by the Poisson bivector
\bea
\label{drinfdouble}
B_{\mathcal{D}(G)}=L_{X_\alpha}\otimes
L_{\xi^\alpha}-R_{X_\alpha}\otimes R_{\xi^\alpha}
\eea
where $L_{X_\alpha}$, $R_{X_\alpha}$, $L_{\xi^\alpha}$ and
$R_{\xi^\alpha}$ denote the right and left invariant vector fields on
$\mathcal{D}(G)$ associated to the basis $X_\alpha$ and its
dual $\xi^\alpha$.

\subsection*{Classical r-matrices, factorising and the Heisenberg
  double structure}
The element $r$ in \eqref{rcocomm} has the properties of a {classical
  $r$-matrix}. A {\em classical
  $r$-matrix}  for a Lie algebra $\frg$ is a tensor
  $r\in\frg\otimes\frg$, whose symmetric part $r+\sigma(r)$ is
   non-degenerate and Ad-invariant and that satisfies the the classical Yang-Baxter
  equation (CYBE)
\bea
\label{cybe}
 [[r,r]]:=[r_{12},r_{13}]+[r_{12},r_{23}]+[r_{13},r_{23}]=0,
\eea
where  $r_{12}=r\otimes r\otimes 1$, $r_{13}=r\otimes 1\otimes r$,
$r_{23}=1\otimes r\otimes r$ and $\sigma$ denotes the flip
$\sigma(a\otimes b)=b\otimes a$.

A classical $r$-matrix allows one  to construct an cocommutator for the
Lie algebra $\frg$ via \eqref{rcocomm} and a Poisson-Lie structure for
the Lie group $G$ given in analogy to \eqref{drinfdouble}. In addition
to this, it defines canonical maps between a  Poisson-Lie group and
its dual. Define the Lie algebra homomorphisms
$\rho,\rho_\sigma:\frg^*\rightarrow\frg$
\bea
\label{rhomap}
\langle\rho(\xi),\eta\rangle:=\langle r,
\xi\otimes\eta\rangle\qquad\langle\rho_\sigma(\xi),
\eta\rangle:=-\langle \sigma(r), \xi\otimes\eta
\rangle\qquad\forall\xi,\eta\in\frg^*,
\eea
where $\langle\;,\;\rangle$ denotes the pairing between $\frg$ and
$\frg^*$. They lift to Lie group homomorphisms
$S,S_\sigma:G^*\rightarrow G$, defining a map $Z:G^*\rightarrow G$ with
$Z(g):=S(g)S_\sigma(g)^{-1}$, which is a local diffeomorphism around the
identity. If it is a global diffeomorphism, the Poisson-Lie group $G$ is
called {\em factorisable} and every $g\in G$ has a unique decomposition
\bea
\label{factor}
g=g_+g_-^{-1}\qquad\mbox{with}\qquad
g_+=S(Z^{-1}(g))\quad\mbox{and}\quad g_-=S_\sigma(Z^{-1}(g)).\eea

Using these maps between a Poisson-Lie group and its
dual, we can embed $G$ and $G^*$ into $G\times G$
\begin{align}
\label{embed}
G&\rightarrow G\times G:\qquad h\rightarrow(h,h) & &\forall h\in G\\
G^*&\rightarrow G\times G:\qquad L\rightarrow(S(L),
S_\sigma(L)) & &\forall L\in G^*.\nonumber
\end{align}
In \cite{setishan},  Semenov-Tian-Shansky introduced  a
Poisson structure
 on $G\times G$ which is such that
 these
embeddings are Poisson maps with respect to the Poisson-Lie structures
on $G$ and $G^*$. The Lie group $G\times G$ with this Poisson structure
is called the {\em Heisenberg double} $\cald_+(G)$ of $G$
(note that it is not  a Poisson-Lie group).  Explicitly, the
Poisson structure is given by the  bivector
\begin{align}
\label{heisdoublebivect}
B'_{\mathcal{D}_+(G)}(g)=&\tfrac{1}{2}r^{\beta\alpha}
(R^1_{X_\alpha}\wedge R^1_{X_\beta}+R^2_{X_\alpha}\wedge
R^2_{X_\beta}+L^1_{X_\alpha}\wedge L^1_{X_\beta}+L^2_{X_\alpha}\wedge L^2_{X_\beta})
\\&+r^{\beta\alpha}(R^1_{X_\alpha}\wedge R^2_{X_\beta}+L^1_{X_\alpha}\wedge
L^2_{X_\beta}),
\nonumber
\end{align}
where the indices 1 and 2 refer to the first and second argument of
$G\times G$ and $L_{X_\alpha}$, $R_{X_\alpha}$ again denote the right
and left invariant vector fields associated to the basis $X_\alpha$.
It was shown in   \cite{wu} that the Poisson structure on the
Heisenberg double of a globally factorisable  group is symplectic.

\subsection{The Poisson-Lie group ${\PPoi}$}

\subsubsection*{The Lie bialgebra $iso(2,1)$ as classical double}
\noindent The Lie bialgebra $iso(2,1)$ with generators $P_a$, $J_a$,
$a=0,1,2$, is is the classical double of $so(2,1)$:
Take $so(2,1)$ with generators $J_a$, $a=0,1,2$, and commutator
\bea
[J_a,J_b]=\epsilon_{abc}J^c.
\eea
Define a dual basis $P_a$, $a=0,1,2$, of the space $so(2,1)^*$
with canonical pairing
\bea
\langle J_a,P^b\rangle=\eta_{ab}
\eea
and equip it with the trivial Lie algebra structure
\bea
[P^a,P^b]=0.
\eea
Then, the element
\bea
\label{rmat}
r=J_a\otimes P^a\qquad\text{and its flip}\qquad \sigma(r)=P^a\otimes J_a
\eea are $r$-matrices for $so(2,1)$.
The commutator on the classical double
$iso(2,1)=so(2,1)\oplus so(2,1)^*$ is given by
\bea
[J_a,J_b]=\epsilon_{ab}^{\;\;\:\;c}J_c\qquad[J_a,P^b]=
\epsilon_{a\;\;c}^{\;\;b}P^c\qquad[P^a,P^b]=0,\eea
the cocommutator by
\bea
\delta(J_a)=0\qquad\delta (P^a)=\epsilon^{a}_{\;\;bc}P^b\otimes P^c.\eea
The connected and simply connected Poisson-Lie
group associated to this Lie bialgebra structure is
the universal cover $\PPoi$ of the Poincar\'e group $\Poi$,
with the multiplication law \eqref{groupmult}.

\subsubsection*{The dual of $\PPoi$}
The dual of Lie bialgebra $iso(2,1)$ is the Lie algebra generated by
$P^a$, $J_a$, $a=0,1,2$, with commutator
\bea
[J_a,J_b]=\epsilon_{ab}^{\;\;\;c}J_c \qquad [J_a,P^b]=0\qquad [P^a,P^b]=0,
\eea
and cocommutator
\bea
\delta(J_a)=\epsilon_{ab}^{\;\;\;c}(P^b\otimes J_c-J_c\otimes P^b)
\qquad\delta(P^a)=\epsilon^a_{\;\;bc}P^b\otimes P^c.
\eea
The unique connected and simply connected Poisson-Lie group
$(\PPoi)^*$ associated to this Lie bialgebra
is the direct product $\LLor\times\RR^3$ with
group multiplication
\bea
(u,\ba)(u',\ba')=(uu',\ba+\ba').
\eea

\subsubsection*{Heisenberg double of $\PPoi$}
\noindent The maps \eqref{rhomap} $\rho$,$\rho_\sigma:
iso(2,1)^*\rightarrow iso(2,1)$ from its dual  into $iso(2,1)$ are given by
\begin{align}
\rho(P_a)&=0 & \rho(J_a)&=J_a\\
\rho_\sigma(P_a)&=-P_a & \rho_\sigma(J_a)&=0.
\end{align}
Their covering maps $S$, $S_\sigma: \left(\PPoi\right)^*
\rightarrow \PPoi$ are obtained using the exponential map
\bea
S(u,\ba)=(u,0) \qquad S_\sigma(u,\ba)=(1,-\Ad(u^{-1})\ba)=(1,\bj).
\eea
They allow to factorise every element $g\in \PPoi$
according to \eqref{factor} as
\bea
\label{factoring}
g=(u,\ba)=(u,\ba)_+(u,\ba)_-^{-1}\eea with
\bea
\label{factel}
(u,\ba)_+=(u,0)\qquad\mbox{and}\qquad(u,\ba)_-=(1,-\Ad(u^{-1})\ba)=(1,\bj).
\eea
Denoting the first argument in $\PPoi\times \PPoi$ by $A$, the second by $B$
and using  the $r$-matrix $r$ in \eqref{rmat}, the Poisson
bivector  \eqref{heisdoublebivect}  on the Heisenberg double of $\PPoi$ becomes
\bea
B'_{\mathcal{D}_+(\PPoi)}(g)&=&\tfrac{1}{2}(P^{A_R}_a
\wedge J^a_{A_R}+P^{B_R}_a\wedge J^a_{B_R}+P^{A_L}_a
\wedge J^a_{A_L}+P^{B_L}_a\wedge J^a_{B_L})\nonumber\\
&+&(P^{A_R}_a\wedge J^a_{B_R}+P^{A_L}_a\wedge J^a_{B_L}).
\eea
After inserting the expressions \eqref{vecfields} for the right and left
invariant vector fields associated to the two arguments
\begin{align}
P_a^{X_R} f(g_A,g_B)&=-\frac{\partial f}{\partial j_X^a}(g_A,g_B)\\
P_a^{X_L}f(g_A,g_B)&=\Ad{(u_X)}_{ab}\frac{\partial f}
{\partial j^X_b}(g_A,g_B)\nonumber\\
J_a^{X_R} f(g_A,g_B)&=\big(\frac{1}{\Ad(u_X)-1}\big)_{ab}\epsilon^b_{cd}p_X^c\frac{\partial f}{\partial p_X^d}(g_A,g_B)-\epsilon_{abc}j_X^b\frac{\partial f}{\partial j^X_c}(g_A,g_B)\nonumber\\
J_a^{X_L} f(g_A,g_B)&=\big(\frac{\Ad(u_X)}{1-\Ad(u_X)}\big)_{ab}
\epsilon_{bcd}p_X^c\frac{\partial f}{\partial p_X^d}(g_A,g_B)
\qquad\text{for}\quad X=A,B\nonumber
\end{align}
and parametrising $g_A$ and $g_B$ as in \eqref{lparam}, we obtain the Poisson bracket on $\mathcal{D}_+(\PPoi)$:
\begin{align}
\label{hdbracket}
&\{j_X^a,p_X^b\}=-\left(\frac{1}{1-\Ad({u_{X}})}\right)^a_{\;\,d}
\epsilon^{bcd}p_c^X
 &
&X=A,B\nonumber\\
&\{j_X^a,j_X^b\}=-\epsilon^{abc}j_c^X &
&X=A,B\nonumber\\
&\{j_A^a,p_B^b\}=-\left(\frac{1}{1-\Ad({u_B})}+\Ad({u^{-1}_A})
\frac{\Ad({u_B})}{1-\Ad({u_{B}})}\right)^a_{\;\;d}\epsilon^{bcd}p_c^B
\nonumber\\
&\{j_A^a,j_B^b\}=-\epsilon^{abc}j_c^B\nonumber\\
&\{j_B^a,p_A^b\}=0\nonumber\\
&\{p_B^a,p_A^b\}=0.
\end{align}

The
Poisson structure on  the Heisenberg double of $\PPoi$ is symplectic.
We give a short, direct proof here.

\begin{theorem}
The Poisson structure \eqref{hdbracket} on the Heisenberg double
of the Poisson-Lie group $\PPoi$ is symplectic.
\end{theorem}

\noindent{\bf Proof:} We have to show that the determinant of
the Matrix representing the Poisson structure \eqref{hdbracket}
with respect to a basis does not vanish. With respect to the (ordered)
basis $B=(\frac{\partial}{\partial p_A^a},\frac{\partial}{\partial
  j_A^a},\frac{\partial}{\partial p_B^a},\frac{\partial}{\partial
  j_B^a})$, $a=0,1,2$,
the Poisson bracket \eqref{hdbracket} is represented by the matrix
\bea
M=\left(\begin{array}{cccc}
0 & -T_A^{-1} & 0 & 0\\
(T_A^{-1})^T & -\epsilon^A & -S & -\epsilon^B\\
0 & S^T & 0 & -T_B^{-1}\\
0 & -\epsilon^B & (T_B^{-1})^T & -\epsilon^B
\end{array}\right)\nonumber
\eea
with
\begin{align} &S^{ab}=\left(\frac{1}{1-\Ad({u_B})}+
\Ad({u^{-1}_A})\frac{\Ad({u_B})}{1-\Ad({u_{B}})}\right)^a_{\;\;d}
\epsilon^{bcd}p_c^B\nonumber\\
&\left(\epsilon^X\right)^{ab}=\epsilon^{abc}j^X_c\nonumber\\
&\left(T_X^{-1}\right)^{ab}=\left(\frac{1}{1-\Ad({u_{X}})}
\right)^a_{\;\,d}\epsilon^{bcd}p_c^X & &\text{for}\;\;X=A,B\;.\nonumber
\nonumber
\end{align}
The determinant of this matrix is given by
\bea
\det M=\left(\det T_A^{-1}\right)^2\left(\det T_B^{-1}\right)^2\nonumber.
\eea
Comparing the matrix $T_X^{-1}$ with the transformation
\eqref{Ttrans} in Sect.~\ref{oneparticle}, which relates the two
sets of position and angular momentum coordinates, we find that it is
given by
\bea
 \left(T_X^{-1}\right)_{ab}=\left(T^{-1}(-\bp^X)\right)_{ab}\qquad\text{for}\;X=A,B.
\eea
As the transformation \eqref{Ttrans} is invertible, its determinant
does not vanish and neither does the determinant of $M$, proving that
the Poisson structure of $\cald_+(\PPoi)$ is symplectic.\; $\Box$

\end{document}